\documentclass{CSML}

\def\dOi{12(3:11)2016}
\lmcsheading%
{\dOi}
{1--38}
{}
{}
{Dec.~30, 2015}
{Sep.~22, 2016}
{}

\ACMCCS{[{\bf Theory of computation}]:  Models of
  computation---Concurrency; Semantics and reasoning---Program
  semantics} 

\usepackage{algorithm}
\usepackage{algpseudocode}
\usepackage{amsmath}
\usepackage{amsfonts}
\usepackage{amssymb}
\usepackage{amsthm}
\usepackage{booktabs}
\usepackage{calc}
\usepackage{hyperref}
\usepackage{stmaryrd}
\usepackage{subfig}
\usepackage{tikz-reo}
\usepackage{wrapfig}

\usepackage{sung-abbr}
\usepackage{sung-code}
\usepackage{sung-hyph}
\usepackage{sung-math}
\usepackage{sung-misc}
\usepackage{sung-tikz}

\newcommand{\edit}[1]{#1} 
\newcommand{\editt}[1]{#1} 

\algrenewtext{EndWhile}{\textbf{od}}
\algrenewtext{EndIf}{\textbf{fi}}
\SetSymbolFont{stmry}{bold}{U}{stmry}{m}{n}

\title[Data optimizations for constraint automata]
      {Data optimizations for constraint automata}

\author[S.-S.T.Q.~Jongmans]{Sung-Shik~T.Q.~Jongmans\rsuper a}
\address{{\lsuper a}Open University of the Netherlands, Radboud University Nijmegen,
	the Netherlands}
\email{ssj@ou.nl}

\author[F.~Arbab]{Farhad Arbab\rsuper b}
\address{{\lsuper b}Centrum Wiskunde \& Informatica, Leiden University,
	the Netherlands}
\email{farhad@cwi.nl}


\begin{document}
	\begin{abstract}
		Constraint automata (CA) \editt{constitute} a coordination model based on finite automata on infinite words.
		Originally introduced for \emph{modeling} of coordinators, an interesting new application of CAs is \emph{implementing} coordinators (i.e., compiling CAs into executable code).
		Such an approach guarantees correctness-by-construction and can even yield code that outperforms hand-crafted code.
		The extent to which these two potential advantages \editt{materialize} depends on the smartness of CA-compilers and the existence of proofs of their correctness.
		
		Every transition in a CA is labeled by a ``data constraint'' that specifies an atomic data-flow between coordinated processes as a first-order formula.
		At run-time, \compilergenerated code must handle data constraints as efficiently as possible.
		In this paper, we present, and prove the correctness of two optimization techniques for CA-compilers related to handling \editt{of} data constraints: a reduction to eliminate redundant variables and a translation from (declarative) data constraints to (imperative) data commands \editt{expressed} in a \edit{small} sequential language.
		Through experiments, we show that these optimization techniques can have a positive impact on performance \editt{of generated executable code}.
	\end{abstract}
	
	\maketitle
	

%
\section{Introduction}
\label{sect:intr}
\subsection*{Context}

In the early 2000s, hardware manufacturers shifted their attention from manufacturing fast\-er---yet purely se\-quen\-tial---uni\-core processors to manufacturing slow\-er---yet increasingly par\-al\-lel---mul\-ti\-core processors.
In the wake of this shift, \emph{concurrent programming} became essential for writing scalable programs on general hardware.
Conceptually, concurrent programs consist of \emph{processes}, which implement modules of sequential computation, and \emph{protocols}, which implement the rules of concurrent interaction that processes must abide by.
As programmers have been writing sequential code for decades, programming processes poses no new fundamental challenges.
What \emph{is} new---and notoriously dif\-fi\-cult---is programming protocols.

In ongoing work, we study an approach to concurrent programming based on syntactic separation of processes from protocols.
In this approach, programmers write their processes in a \emph{general-purpose language} (\GPL), while they write their protocols in a complementary \emph{domain-specific language} (\DSL).
Paraphrasing the definition of \DSLs by Van Deursen et al.~\cite{vDKV00}, a \DSL for protocols ``is a programming language that offers, through appropriate notations and abstractions, expressive power focused on, and [..] restricted to, [programming protocols].''
In developing \DSLs for protocols, we draw inspiration from existing \emph{coordination models} and \emph{languages}, which 
typically provide high-level constructs and abstractions that more easily compose into correct---relative to programmers' intentions---protocol code than do lower-level synchronization mechanisms (e.g., locks or sem\-a\-phores).
Significant as their software engineering advantages may be, however, performance is an important concern too.
A crucial step toward adoption of coordination models and languages for programming protocols is, therefore, the development of compilers capable of generating efficient lower-level protocol implementations from high-level protocol specifications.

Our current work focuses on developing compilation technology for \emph{constraint automata} (\CA)~\cite{BSAR06,Jon16}, a coordination model based on finite automata on infinite words, originating from research on the coordination language Reo~\cite{Arb04,Arb11}.
Every \CA models (the behavior of) a \emph{coordinator} that enforces a protocol among coordinated processes.
Structurally, a \CA consists of a finite set of states, a finite set of transitions, a set of directed \emph{ports}, and a set of local \emph{memory cells}.
States model the internal configurations of a coordinator; transitions model a coordinator's atomic coordination steps.
Ports constitute the interface between a coordinator and its coordinated processes, the latter of which can perform blocking \IO-operations on their coordinator's ports: a coordinator's \emph{input ports} admit \code{put} operations, while its \emph{output ports} admit \code{get} operations.
Memory cells model a coordinator's internal buffers to temporarily store data in.
Different from classical automata, transition labels of \CAs consist of two elements: a set of ports, called a \emph{synchronization constraint}, and a logical formula over ports and memory cells, called a \emph{data constraint}.
A synchronization constraint specifies which ports need an \IO-operation for its transition to fire (i.e., those ports synchronize in that transition and their pending \IO-operations complete), while a data constraint specifies which particular data those \IO-operations may involve.
Every \CA, then, constrains \emph{when} \IO-operations may complete on \emph{which} ports.

\subsection*{Problem}

Briefly, our current \CA-to-Java compiler translates passive data structures for \CAs into (re)active ``coordinator threads''.
A coordinator thread is, effectively, a state machine whose \edit{transitions correspond} one-to-one to \edit{transitions} in a \CA.
Essentially, then, \compilergenerated coordinator threads simulate \CAs by firing their transitions, continuously monitoring run-time data structures for their ports.\footnote{%
	One needs to overcome a number of serious issues before this approach can yield practically useful code.
	Most significantly, these issues include exponential explosion of the number of states or transitions of \CA, and \emph{oversequentialization} \edit{(i.e., the situation where coordinator threads \editt{unnecessarily prevent concurrent execution of independent activities})} or \emph{overparallelization} \edit{(i.e., the situation where the synchronization necessary for parallel execution of multiple coordinator threads dominates execution time, to the extent that concurrency causes slowdown instead of speedup)} of generated code.
	We have already reported our work on these issues along with promising results elsewhere~\cite{JA16,JHA14a,JSA15}.
}

To actually fire a transition, a coordinator thread must first check both that transition's synchronization constraint and its data constraint.
The check for the synchronization constraint ensures that all ports involved in the transition have a pending \IO-operation (and are thus ready to participate in the transition); the check for the data constraint subsequently ensures that those pending \IO-operations can result in admissible data-flows.

Checking synchronization constraints is relatively cheap.
Checking data constraints, in contrast, requires calls to a constraint solver.
When using a general-purpose constraint solver, as we \edit{currently} do, such calls inflict high run-time overhead.
This overhead has a severe impact on the overall performance of programs, because coordinator threads execute purely serially.
As such, checking data constraints can become a serious sequential bottleneck to an entire program (e.g., whenever all other threads depend on the firing of a transition to make progress).

\subsection*{Contribution}

In this paper, we present two techniques to optimize the performance of checking data constraints.
The first technique reduces the size of data constraints at compile-time, to reduce the complexity \editt{of} (number of variables \editt{involved in}) constraint solving at run-time.
The second technique translates data constraints into \editt{small} pieces of imperative code (in a sequential language with assignment and guarded failure \editt{statements}) at compile-time, to replace expensive calls to a general-purpose constraint solver at run-time.
We prove that both our techniques are correct.
Such correctness proofs are important, because they ensure that our compilation approach guarantees \emph{correctness-by-construction} (e.g., model-checking results obtained for pre-optimized \CA also hold for their generated, optimized implementations).
We evaluate our techniques in a number of experiments using their implementation in our current \CA-to-Java compiler.

In Section~\ref{sect:prel}, we present preliminaries on data constraints and \CAs.
In Section~\ref{sect:elim}, we discuss our first optimization technique; in Section~\ref{sect:comm}, we discuss our second.
In Section~\ref{sect:exper}, we report on an experimental evaluation of our two optimization techniques.
Section~\ref{sect:concl} concludes this paper.
Appendix~\ref{sect:proofs} contains proof sketches; full, detailed proofs appear in~\cite{techreport} (referenced more specifically in Appendix~\ref{sect:proofs}).
A preliminary version of this paper, in which we report only on the
optimization technique presented in Section~\ref{sect:comm}, appeared
in the proceedings of \textsc{Coordination} 2015~\cite{JA15b}.

%
\section{Preliminaries}
\label{sect:prel}
\subsection*{Data Constraints}

In this subsection, we present a first-order calculus of data constraints.
In the next subsection, \edit{we label transitions in \CAs with objects from this calculus.}
We start by defining elementary notions of data, ports, and memory cells.

\begin{defi}
	[data]
	\label{def:duniv}
	A datum is an unstructured object.
	$\duniv$ denotes the \editt{possibly infinite} set of all data, ranged over by~$d$.
\end{defi}

\begin{defi}
	[empty datum]
	\label{def:nil}
	$\nil$ is an unstructured object such that~$\nil \notin \duniv$.
\end{defi}

\begin{defi}
	[ports]
	\label{def:puniv}
	A port is an unstructured object.
	$\puniv$ denotes the set of all ports, ranged over by~$p$.
	$\pset{\puniv}$ denotes the set of all sets of ports, ranged over by~$P$.
\end{defi}

\begin{defi}
	[memory cells]
	\label{def:muniv}
	A memory cell is an unstructured object.
	$\muniv$ denotes the set of all memory cells, ranged over by~$m$.
	$\pset{\muniv}$ denotes the set of all sets of memory cells, ranged over by~$M$.
\end{defi}

The exact content of~$\duniv$ depends on the context of its use and formally does not matter.
Henceforth, we write elements of~$\puniv$ in capitalized lower case sans-serif (e.g., \reo{A}, \reo{B}, \reo{C}, \reo{In}$_1$, \reo{Out}$_2$), while we write elements of~$\duniv$ in lower case monospace (e.g., \code{1}, \code{3.14}, \code{true}, \code{"foo"}).
Although data flow through ports always in a certain direction, we do not yet distinguish input ports from output ports; this comes later.

Out of ports and memory cells, we construct \emph{data variables}, which serve as the variables in our calculus.
Every data variable \editt{designates a datum}.
For instance, ports can hold data \editt{(to exchange)}, so every port serves as a data variable in the calculus.
Similarly, memory cells can hold data, but the meaning of ``to hold'' differs in this case.
Ports hold data only \editt{for exchange} \emph{during} a coordination step (i.e., transiently, in passing).
In contrast, memory cells hold data also \emph{before} and \emph{after} a coordination step.
Consequently, in the context of data variables, a memory cell before a coordination step and the same memory cell after that step have different \editt{identities}.
After all, the content of the memory cell may have changed in between.
Therefore---inspired by notation from Petri nets~\cite{Rei85}---for every memory cell~$m$, both~$\pre{m}$ and~$\post{m}$ serve as data variables:~$\pre{m}$ refers to the datum in~$m$ before a coordination step, while~$\post{m}$ refers to the datum in~$m$ after that coordination step.
We abbreviate sets~$\setb{\pre{m}}{m \in M}$ and~$\setb{\post{m}}{m \in M}$ as~$\pre{M}$ and~$\post{M}$.

\begin{defi}
	[data variables]
	\label{def:xuniv}
	A data variable is an object~$x$ generated by the following grammar:
	$$
	\begin{array}{@{} l @{\ } c @{\ } l @{}}
		x	& ::=	& p \mid \pre{m} \mid \post{m}
	\end{array}
	$$
	$\xuniv$ denotes the set of all data variables.
	$\pset{\xuniv}$ denotes the set of all sets of data variables, ranged over by~$X$.
\end{defi}

We subsequently assign meaning to data variables with \emph{data assignments}.

\begin{defi}
	[data assignments]
	\label{def:assignmuniv}
	A data assignment is a partial function from data variables to data.
	$\assignmuniv = \xuniv \rightharpoonup \duniv$ denotes the set of all data assignments, ranged over by~$\sigma$.
	$\pset{\assignmuniv}$ denotes the set of all sets of data assignments, ranged over by~$\Sigma$.
\end{defi}

\noindent
Essentially, a data assignment~$\sigma$ comprehensively models a coordination step involving the ports and memory cells in~$\dom{\sigma}$ and the data in~$\img{\sigma}$.
\edit{As coordinators have only finitely many ports and memory cells in practice, we stipulate that the domain of every data assignment is finite, too.
The same holds for their support.}

We proceed by defining \emph{data functions} and \emph{data relations}, which serve as the functions and predicates in our calculus.
Together, data, data functions, and data relations constitute our set of extralogicals.
To avoid excessive machinery---but at the cost of formal \editt{imprecision}---we do not distinguish extralogical symbols from their interpretation as data, data functions, and data relations.

\begin{defi}
	[data functions]
	\label{def:funiv}
	A data function is a function from tuples of data to data.
	$\funiv = \bigcup \setb{\duniv^k \rightarrow \duniv}{k > 0}$ denotes the set of all data functions, ranged over by~$f$.
\end{defi}

\begin{defi}
	[data relations]
	\label{def:runiv}
	A data relation is a relation on tuples of data.
	$\runiv = \bigcup \setb{\pset{\duniv^k}}{k > 0}$ denotes the set of all data relations, ranged over by~$R$.
\end{defi}

Henceforth, we write elements of~$\funiv$ in camel case monospace (e.g., \code{divByThree}, \code{inc}), while we write elements of~$\runiv$ in captitalized camel case monospace (e.g., \code{Odd}, \code{SmallerThan}).

Out of data variables, data, and data functions, we construct \emph{data terms}, which serve as the terms in our calculus.
Every data term represents a datum.

\begin{defi}
	[data terms]
	\label{def:termuniv}
	A data term is an object~$t$ generated by the following grammar:
	$$
	\begin{array}{@{} l @{\ } c @{\ } l @{}}
		t	& ::=	& x \mid d \mid f(t_1 \, \ldots \, t_{k\geq1})
	\end{array}
	$$
	$\termuniv$ denotes the set of all data terms.
	$2^\termuniv$ denotes the set of all sets of data terms, ranged over by~$T$.
\end{defi}

\noindent
Henceforth, let~$\termorder$\index{<termorder@$\termorder$} denote some strict total order on~$\termuniv$.\footnote{%
	\edit{It does not matter what this strict total order exactly looks like, so long as we have some way of selecting the least element of any set of terms.
	We use this property in Definition~\ref{def:existsfun}.}
}

Given a data assignment whose domain includes at least the data variables in a data term~$t$, we can \emph{evaluate}~$t$ to a datum.
(To evaluate~$t$, additionally, every data function application in~$t$ must have the right number of inputs: the \emph{arity} of a data function and its number of inputs must match.
Henceforth, we tacitly assume that this always holds true.)

\begin{defi}
	[evaluation]
	\label{def:evalfun}
	$\evalfun : \assignmuniv \times \termuniv \rightarrow \duniv \cup \set{\nil}$ denotes the function defined by the following equations:
	$$
	\begin{array}{@{} l @{\ } c @{\ } l @{}}
		\evalfun_\sigma(x)							& =	& \left \{ \begin{array}{@{} l @{} l @{}}
															\sigma(x)	& \IF x \in \dom{\sigma}
														\\	\nil		& \OTHERWISE
														\end{array} \right.
	\\	\UP
	\\	\evalfun_\sigma(d)							& =	& d
	\\	\UP
	\\	\evalfun_\sigma(f(t_1 \, \ldots \, t_k))	& =	& \left \{ \begin{array}{@{} l @{} l @{}}
															f(\evalfun_\sigma(t_1) \, \ldots \, \evalfun_\sigma(t_k))	& \IF \SCOPEX{%
																															\evalfun_\sigma(t_1) \neq \nil
																														\\	\LAND \cdots \RAND
																														\\	\evalfun_\sigma(t_k) \neq \nil
																														}
														\\	\UP
														\\	\nil														& \OTHERWISE
														\end{array} \right .
	\end{array}
	$$
\end{defi}

\noindent Out of data terms, data relations, and data variables, we construct data constraints.

\begin{defi}
	[data constraints]
	\label{def:dcuniv}
	A data constraint is an object~$\varphi$ generated by the following grammar:
	$$
	\begin{array}{@{} l @{\enspace} c @{\enspace} l @{\qquad} l @{}}
		a		& ::=	& \bot \mid \top \mid \edit{t_1} \termeq \edit{t_2} \mid R(t_1 \, \ldots \, t_{k\geq1})									& \text{(data atoms)}
	\\	\ell	& ::=	& a \mid \neg a	& \text{(data literals)}
	\\	\varphi	& ::=	& \exists x . \varphi \mid \ell_1 \wedge \cdots \wedge \ell_{k\geq1}	& \text{(data constraints)}
	\end{array}
	$$
	$\dcuniv$ denotes the set of all data constraints.
	$\pset{\dcuniv}$ denotes the set of all sets of data constraints, ranged over by~$\Phi$.
\end{defi}

Henceforth, let~$\dcorder$ denote a strict total order on~$\dcuniv$, and let~$\dcbigand \Phi$ denote the \emph{unique} multiary conjunction of the data constraints in~$\Phi$ under~$\dcorder$.
Also, for a data constraint $\varphi = \exists x_1 . \cdots \exists x_l . (\ell_1 \wedge \cdots \wedge \ell_k)$, call $\ell_1 \wedge \cdots \wedge \ell_k$ the \emph{kernel} of $\varphi$, and let $\litfun(\varphi) = \set{\ell_1 , \ldots , \ell_k}$.

Every data constraint characterizes a set of data assignments through an \emph{entailment relation}.
\edit{This entailment relation, thus, formalizes the semantics of data constraints.}
Let~$\varphi [t / x]$ denote data constraint~$\varphi$ with data term~$t$ substituted for every occurrence of data variable~$x$ (in a capture-free way).

\begin{figure}
	\centering
	
	\renewcommand{\WIDTH}{.5\linewidth}
	\begin{framed}
		\begin{tabular}{@{}c@{}c@{}}
			\begin{minipage}{\WIDTH}
				\noindent%
				\begin{equation}
					\hphantom{xxxxxxx} \dfrac{%
						\strut%
					}{%
						\sigma \dcmodels \top
					} \hphantom{xxxxxxx}
				\end{equation}
			\end{minipage}
		&	\begin{minipage}{\WIDTH}
				\noindent%
				\begin{equation}
					\dfracx[l]{%
						\strut%
						\freefun(a) \subseteq \dom{\sigma} \AND \sigma \not \dcmodels a 
					}{%
						\sigma \dcmodels \neg a
					}
				\end{equation}
			\end{minipage}
		\\
		\\	\begin{minipage}{\WIDTH}
				\noindent%
				\begin{equation}
					\dfrac{%
						\strut%
						\evalfun_\sigma(t_1) = \evalfun_\sigma(t_2) \neq \nil
					}{%
						\sigma \dcmodels t_1 \termeq t_2
					}
				\end{equation}
			\end{minipage}
		&	\begin{minipage}{\WIDTH}
				\noindent%
				\begin{equation}
					\dfracx{%
						\strut%
						\sigma \dcmodels \phi[d / x] \FORSOME d
					}{%
						\sigma \dcmodels \exists x . \phi
					}
				\end{equation}
			\end{minipage}
		\\
		\\	\begin{minipage}{\WIDTH}
				\noindent%
				\begin{equation}
					\dfrac{%
						\strut%
						\tpl{\evalfun_\sigma(t_1) \, \ldots \, \evalfun_\sigma(t_k)} \in R
					}{%
						\sigma \dcmodels R(t_1 \, \ldots \, t_k)
					}
				\end{equation}
			\end{minipage}
		&	\begin{minipage}{\WIDTH}
				\noindent%
				\begin{equation}
					\dfrac{%
						\strut%
						\sigma \dcmodels \phi_1 \AND \cdots \AND \sigma \dcmodels \phi_k
					}{%
						\sigma \dcmodels \phi_1 \wedge \cdots \wedge \phi_k
					}
				\end{equation}
			\end{minipage}
		\end{tabular}
	\end{framed}
	
	\caption{Addendum to Definition~\ref{def:dcmodels}}
	\label{fig:def+dcmodels}
\end{figure}

\begin{defi}
	[entailment]
	\label{def:dcmodels}
	${\dcmodels} \subseteq \assignmuniv \times \dcuniv$ denotes the smallest relation induced by the rules in Figure~\ref{fig:def+dcmodels}.
\end{defi}

Contradiction, tautology, and (multiary) conjunction have standard semantics~\cite{Rau10a}.
Negation~$\neg a$ means that, despite all free variables in~$a$ having a value,~$a$ does not hold true; the extra condition on the free variables in~$a$ ensures the \emph{monotonicity} of entailment (i.e.,~$\sigma |_X \dcmodels \varphi$ implies~$\sigma \dcmodels \varphi$, for all~$X , \varphi$).
Data atom~$t_1 \termeq t_2$ means that~$t_1$ and~$t_2$ evaluate to the same datum.
Typical examples include~$p_1 \termeq p_2$ (i.e., the same datum passes through ports~$p_1$ and~$p_2$),~$p \termeq \post{m}$ (i.e., the datum that passes through port~$p$ enters the buffer modeled by memory cell~$m$), and~$p \termeq \pre{m}$ (i.e., the datum in the buffer modeled by memory cell~$m$ exits that buffer and passes through port~$p$).
Tautology~$\top$ means that it does not matter which data flow through which ports.

Henceforth, let~$\dcimpl$ and~$\dcequiv$ denote the implication relation and the equivalence relation on data constraints, derived from~$\dcmodels$ in the usual way~\cite{Rau10a}.
Furthermore, let~$\variablfun(\varphi)$ denote the set of data variables in~$\varphi$, and let~$\freefun(\varphi)$ denote its set of \emph{free} data variables.

\subsection*{Constraint Automata}

We proceed by formally defining a \CA~$\bfa$, which models a coordinator, as a tuple consisting of a set of states~$Q$, a triple of three sets of ports~$\tpl{P^\textall \, P^\textin \, P^\textout}$, a set of memory cells~$M$, a transition relation~$\longrightarrow$, and an initial state~$q^0$.
\editt{The} set~$P^\textall$ contains all ports monitored and controlled by~$\bfa$, while~$P^\textin$ and~$P^\textout$ contain only its input ports and its output ports.
Although~$P^\textall$ contains the union of~$P^\textin$ and~$P^\textout$, the converse not necessarily holds true: beside input and output ports,~$P^\textall$ may contain also \emph{internal ports}\index{port!internal}.
If a \CA has internal ports, we call it a \emph{composite}; otherwise, we call it a \emph{primitive}.

\begin{defi}
	[states]
	\label{def:quniv}
	A state \editt{represents a configuration of a coordinator}.
	$\quniv$ denotes the set of all states, ranged over by~$q$.
	$\pset{\quniv}$ denotes the set of all sets of states, ranged over by~$Q$.
\end{defi}

\begin{defi}
	[constraint automata]
	\label{def:automuniv}
	A constraint automaton is a tuple:
	$$
	\tpl{Q \, \tpl{P^\textall \, P^\textin \, P^\textout} \, M \, \longrightarrow \, q^0}
	$$
	where:
	\begin{itemize}
		\item
		$Q \subseteq \quniv$ \hfill (states)
		
		\item
		$\tpl{P^\textall \, P^\textin \, P^\textout} \in \pset{\puniv} \times \pset{\puniv} \times \pset{\puniv}$ such that: \hfill (ports)
		$$
		P^\textin \, P^\textout \subseteq P^\textall \AND P^\textin \cap P^\textout = \emptyset
		$$
		
		\item
		$M \subseteq \muniv$ \hfill (memory cells)
		
		\item
		${\longrightarrow} \subseteq Q \times \pset{P^\textall} \times \dcuniv \times Q$ such that: \hfill (transitions)
		$$
		\SCOPE{q \xrightarrow{P,\varphi} q' \IMPLIES \freefun(\varphi) \subseteq P \cup \pre{M} \cup \post{M}} \FORALL q \, q' \, P \, \varphi
		$$
		
		\item
		$q^0 \in Q$ \hfill (initial state)
	\end{itemize}
	$\automuniv$ denotes the set of all constraint automata, ranged over by~$\bfa$.
\end{defi}

\begin{figure}[t]
	\captionsetup[subfloat]{labelformat=empty}%
	\subfloatvspace
	\hfil
	\subfloatv[104pt]{Textual representation}{%
		$\begin{array}{@{} l @{\ } c @{\ } l @{}}
			Q											& =	& \set{q_1 \, q_2}
		\\	\tpl{P^\textall \, P^\textin \, P^\textout}	& =	& \tpl{\set{\reo{A} \, \reo{B} \, \reo{C}} \, \set{\reo{A} \, \reo{B}} \, \set{\reo{C}}}
		\\	M											& =	& \set{\reo{x}}
		\\	\UP
		\\	{\longrightarrow}							& =	& \setx{%
																\tpl{q_1 \, \set{\reo{A}} \, \reo{A} \termeq \post{\reo{x}} \, q_2} \,[]
															\\	\tpl{q_1 \, \set{\reo{B}} \, \reo{B} \termeq \post{\reo{x}} \, q_2} \,[]
															\\	\tpl{q_2 \, \set{\reo{C}} \, \pre{\reo{x}} \termeq \reo{C} \, q_1} \,[]
															}
		\\	\UP
		\\	q^0											& =	& q_1
		\end{array}$
	}
	\hfil
	\subfloatv[104pt]{Graphical representation}{%
		\renewcommand{\reosize}{\tiny}%
		\qquad \begin{tikzautom}
			\state{Q1}{0,0}{}
			\state{Q2}{right of=Q1,node distance=2*\dist cm}{}
			\init{left of}{Q1}
			\trans{Q1}{Q2}{from16,to9}{below}{\set{\reo{A}^\textin} \, \reo{A} \termeq \post{\reo{x}}}
			\trans{Q1}{Q2}{from14,to11}{below}{\set{\reo{B}^\textin} \, \reo{B} \termeq \post{\reo{x}}}
			\trans{Q2}{Q1}{from8,to1}{above}{\set{\reo{C}^\textout} \, \pre{\reo{x}} \termeq \reo{C}}
		\end{tikzautom} \qquad
	}
	\hfil
	
	\caption{Example \CA for a producers\slash consumer coordinator}
	\label{fig:autom+lateasyncmerger2}
\end{figure}

\noindent
The requirement~$\freefun(\varphi) \subseteq P \cup \pre{M} \cup \post{M}$ means that the effect of a transition remains local to its own scope: a transition cannot affect, or be affected by, ports outside its synchronization constraint and memory cells outside its \CA.
Henceforth, let~$\dcfun(\bfa)$ denote the set of data constraints that occur on the transitions of a \CA~$\bfa$ \edit{(not to be confused with~$\dcuniv$, which denotes the set of all data constraints; see Definition~\ref{def:dcuniv})}.

Figure~\ref{fig:autom+lateasyncmerger2} shows an example of a \CA.
In graphical representations of \CAs, we annotate ports in synchronization constraints with superscripts ``in'' and ``out'' to indicate their direction; internal ports have no such annotation.
The \CA in Figure~\ref{fig:autom+lateasyncmerger2} models a producers\slash consumer coordinator with two input ports \reo{A} and \reo{B} (each shared with a different producer, presumably) and an output port \reo{C} (shared with the consumer).
Initially, a \code{put} by the producer on \reo{A} can complete, causing that producer to offer a datum into internal buffer \reo{x} (modeled by data constraint~$\reo{A} \termeq \post{\reo{x}}$).
Alternatively, a \code{put} by the other producer on \reo{B} can similarly complete.
Subsequently, only a \code{get} by the consumer on \reo{C} can complete, causing the consumer to accept the datum previously stored in \reo{x}.
This coordinator, thus, enforces asynchronous, unordered, reliable communication from two producers to a consumer.

The precise definitions of \emph{language acceptance} and \emph{bisimulation} for \CAs do not matter in this paper.
Likewise, the precise definitions of \emph{behavioral equivalence} (based on language acceptance) and \emph{behavioral congruence} (based on bisimulation), such that behavioral congruence implies behavioral equivalence, do not matter.
These definitions appear elsewhere~\cite{Jon16}.
The only result about the behavior of \CAs that matters in this paper is the following intuitive proposition.
Let~$\congr$ denote behavioral congruence, and let~${\bfa} [\varphi' / \varphi]$ denote \CA~$\bfa$ with data constraint~$\varphi'$ substituted for every occurrence of data constraint~$\varphi$.

\begin{prop}
	[Lemma 43 in {\cite[Appendix C.4]{techreport}}]
	\label{prop:001}
	$\varphi \dcequiv \varphi' \IMPLIES \bfa \congr {\bfa} [\varphi' / \varphi]$
\end{prop}

This proposition means that we can freely replace every data constraint in a \CA with an equivalent data constraint in a behavior-neutral way.
This proposition plays a key role in the correctness proofs of the two optimization techniques presented in the rest of this paper.

\begin{figure}[t]
	\captionsetup[subfloat]{labelformat=empty}
	\subfloatvspace
	\renewcommand{\HEIGHT}{57pt}
	\renewcommand{\reosize}{\scriptsize}
	\hfil
	\subfloatv[\HEIGHT]{{}\protect\reoname{Sync}{}{}{p_1 ; p_2}}{\hphantom{x} \tikzautomone{}{}{\set{p_1^\textin \, p_2^\textout} \, p_1 \termeq p_2}{}{init}{}{}{} \hphantom{x}}
	\hfil
	\subfloatv[\HEIGHT]{{}\protect\reoname{SyncDrain}{}{}{p_1 , p_2 ;}}{\hphantom{xxxxx} \tikzautomone{}{}{\set{p_1^\textin \, p_2^\textin} \, \top}{}{init}{}{}{} \hphantom{xxxxx}}
	\hfil
	\subfloatv[\HEIGHT]{{}\protect\reoname{LossySync}{}{}{p_1 ; p_2}}{\hphantom{xx} \tikzautomone{}{\set{p_1^\textin} \, \top}{}{}{init}{}{}{\set{p_1^\textin \, p_2^\textout} \,[] \\ p_1 \termeq p_2} \hphantom{xx}}
	\hfil
	\subfloatv[\HEIGHT]{{}\protect\reoname{Filter}{R}{}{p_1 ; p_2}}{\tikzautomone{}{\set{p_1^\textin} \, \neg R(p_1)}{}{}{init}{}{}{\set{p_1^\textin \, p_2^\textout} \,[] \\ R(p_1) \wedge p_1 \termeq p_2}}
	\hfil
	
	\renewcommand{\HEIGHT}{57pt}
	\hfil%
	\subfloatv[\HEIGHT]{{}\protect\reoname{Fifo}{}{;m}{p_1 ; p_2}}{%
		\hphantom{xxx}\begin{tikzautom}
			\state{Q1}{0,0}{};
			\state{Q2}{right of=Q1,node distance=1*\dist cm}{};
			\init{left of}{Q1};
			\trans{Q1}{Q2}{from16,to9}{below}{\set{p_1^\textin} \,[] \\ \post{m} \termeq p_1};
			\trans{Q2}{Q1}{from8,to1}{above}{\set{p_2^\textout} \,[] \\ p_2 \termeq \pre{m}};
		\end{tikzautom}\hphantom{xxx}%
	}%
	\hfil%
	\subfloatv[\HEIGHT]{{}\protect\reoname{Merg2}{}{}{p_1 , p_2 ; p_3}}{\hphantom{x} \tikzautomone{}{\set{p_1^\textin \, p_3^\textout} \,[] \n p_1 \termeq p_3}{}{}{init}{}{}{\set{p_2^\textin \, p_3^\textout} \,[] \n p_2 \termeq p_3} \hphantom{x}}
	\hfil%
	\subfloatv[\HEIGHT]{{}\protect\reoname{Repl2}{}{}{p_1 ; p_2 , p_3}}{\hphantom{xxx} \tikzautomone{}{}{\set{p_1^\textin \, p_2^\textout \, p_3^\textout} \,[] \n p_1 \termeq p_2 \wedge p_1 \termeq p_3}{}{init}{}{}{} \hphantom{xxx}}
	\hfil%
	\subfloatv[\HEIGHT]{{}\protect\reoname{BinOp}{f}{}{p_1 , p_2 ; p_3}}{\hphantom{xxxx} \tikzautomone{}{}{\set{p_1^\textin \, p_2^\textin \, p_3^\textout} \,[] \n f(p_1 \, p_2) \termeq p_3}{}{init}{}{}{} \hphantom{xxxx}}
	\hfil%
	
	\caption{Eight primitives}
	\label{fig:autom+core}
\end{figure}

\begin{figure}[t]
	\centering
	\renewcommand{\WIDTH}{\linewidth-106pt}
	\begin{tabular}{@{} l @{\quad} l @{}}
		\toprule
		\reoname{Sync}{}{}{p_1 ; p_2}			& \begin{minipage}[t]{\WIDTH}Infinitely often atomically~$\big[$ac\-cepts a datum~$d$ on its input port~$p_1$, then offers~$d$ on its output port~$p_2 \big]$.\end{minipage}
	\\	\UP
	\\	\reoname{SyncDrain}{}{}{p_1 , p_2 ;}	& \begin{minipage}[t]{\WIDTH}Infinitely often atomically \SCOPE{ac\-cepts data~$d_1$ and~$d_2$ on its input ports~$p_1$ and~$p_2$, then loses~$d_1$ and~$d_2$}.\end{minipage}
	\\	\UP
	\\	\reoname{LossySync}{}{}{p_1 ; p_2}		& \begin{minipage}[t]{\WIDTH}Infinitely often either atomically \SCOPE{ac\-cepts a datum~$d$ on its input port~$p_1$, then offers~$d$ on its output port~$p_2$} or atomically \SCOPE{accepts a datum~$d$ on~$p_1$, then loses~$d$}.\end{minipage}
	\\	\UP
	\\	\reoname{Filter}{R}{}{p_1 ; p_2}		& \begin{minipage}[t]{\WIDTH}Infinitely often either atomically \SCOPE{ac\-cepts a datum~$d$ on its input port~$p_1$, then establishes that~$d$ satisfies data relation~$R$, then offers~$d$ on its output port~$p_2$} or atomically \SCOPE{accepts a datum~$d$ on~$p_1$, then establishes that~$d$ violates~$R$, then loses~$d$}.\end{minipage}
	\\	\UP
	\\	\reoname{Fifo}{}{;m}{p_1 ; p_2}			& \begin{minipage}[t]{\WIDTH}Infinitely often first atomically \SCOPE{ac\-cepts a datum~$d$ on its input port~$p_1$, then stores~$d$ in its memory cell~$m$} and subsequently atomically \SCOPE{loads~$d$ from~$m$, then offers~$d$ on its output port~$p_2$}.\end{minipage}
	\\	\UP
	\\	\reoname{Merg2}{}{}{p_1 , p_2 ; p_3}	& \begin{minipage}[t]{\WIDTH}Infinitely often atomically \SCOPE{accepts a datum~$d$ either on its input port~$p_1$ or on its input port~$p_2$, then offers~$d$ on its output port~$p_3$}.\end{minipage}
	\\	\UP
	\\	\reoname{Repl2}{}{}{p_1 ; p_2 , p_3}	& \begin{minipage}[t]{\WIDTH}Infinitely often atomically \SCOPE{ac\-cepts a datum~$d$ on its input port~$p_1$, then offers~$d$ on its output ports~$p_2$ and~$p_3$}.\end{minipage}
	\\	\UP
	\\	\reoname{BinOp}{f}{}{p_1 , p_2 ; p_3}	& \begin{minipage}[t]{\WIDTH}Infinitely often atomically \SCOPE{ac\-cepts data~$d_1$ and~$d_2$ on its input ports~$p_1$ and~$p_2$, then applies data function~$f$ to~$d_1$ and~$d_2$, then offers~$f(d_1 \, d_2)$ on its output port~$p_3$}.\end{minipage}
	\\	\bottomrule
	\end{tabular}
	
	\caption{Data-flow behavior of the primitives in Figure~\ref{fig:autom+core}}
	\label{fig:descr+core}
\end{figure}

Instead of defining \CAs directly, in practice, we construct them \emph{compositionally} using two binary operations~\cite{BSAR06,Jon16}: \emph{join}, denoted by~$\otimes$, and \emph{hide}, denoted by~$\ominus$.
Join performs parallel composition: it ``glues'' together two \CAs on their shared ports, after which those shared ports become internal.
Essentially, whenever two \CAs have joined, if a transition in one of those \CAs involves shared ports, that transition can fire only synchronously with a transition in the other \CA that involves \emph{exactly the same} shared ports (i.e., at any time, the \CAs must agree on firing transitions involving their shared ports).
Hide performs port abstraction: it ``cuts'' a port \editt{out} from a \CA.
Typically, we use hide to remove internal ports from the definition of a \CA, as such ports do not directly contribute to its observable behavior (i.e., processes cannot perform \IO-operations on internal ports).
To compositionally construct a \CA, then, we first join a number of ``small'' primitive \CAs into a ``large'' composite \CA.
Second, we hide all internal ports from this large \CA to make its definition more concise (without losing essential information).
Figure~\ref{fig:autom+core} shows a number of common primitive \CAs; Figure~\ref{fig:descr+core} explains their behavior in terms of data-flows between their ports.
In these figures, every \CA has a signature formatted as follows:
$$
\text{\emph{name}} \langle \text{\emph{extralogicals}} \rangle \{\text{\emph{internal ports\:;\:memory cells}}\} (\text{\emph{input ports\:;\:output ports}})
$$

\begin{figure}[t]
	\captionsetup[subfloat]{labelformat=empty}
	\subfloatvspace
	\renewcommand{\reosize}{\scriptsize}
	
	\renewcommand{\HEIGHT}{20pt}
	\hfil
	\subfloatv[\HEIGHT]{{}\protect\reoname{Sync}{}{}{p_1 ; p_2}}{%
		\hphantom{xxx}\begin{tikzcirc}
			\port{P1}{0,0}{above}{p_1}
			\port{P2}{right of=P1}{above}{p_2}
			\sync{P1}{P2}
		\end{tikzcirc}\hphantom{xxx}%
	}
	\hfil
	\subfloatv[\HEIGHT]{{}\protect\reoname{SyncDrain}{}{}{p_1 , p_2 ;}}{%
		\hphantom{xxxxx}\begin{tikzcirc}
			\port{P1}{0,0}{above}{p_1}
			\port{P2}{right of=P1}{above}{p_2}
			\syncdrain{P1}{P2}
		\end{tikzcirc}\hphantom{xxxxx}%
	}
	\hfil
	\subfloatv[\HEIGHT]{{}\protect\reoname{LossySync}{}{}{p_1 , p_2 ;}}{%
		\hphantom{xxxxx}\begin{tikzcirc}
			\port{P1}{0,0}{above}{p_1}
			\port{P2}{right of=P1}{above}{p_2}
			\lossysync{P1}{P2}
		\end{tikzcirc}\hphantom{xxxxx}%
	}
	\hfil
	\subfloatv[\HEIGHT]{{}\protect\reoname{Filter}{R}{}{p_1 ; p_2}}{%
		\hphantom{xxxx}\begin{tikzcirc}
			\port{P1}{0,0}{above}{p_1}
			\port{P2}{right of=P1}{above}{p_2}
			\filter{P1}{P2}{R}
		\end{tikzcirc}\hphantom{xxxx}%
	}
	\hfil
	
	\renewcommand{\HEIGHT}{39pt}
	\hfil
	\subfloatv[\HEIGHT]{{}\protect\reoname{Fifo}{}{;m}{p_1 ; p_2}}{%
		\hphantom{xxxx}\begin{tikzcirc}
			\port{P1}{0,0}{above}{p_1}
			\port{P2}{right of=P1}{above}{p_2}
			\fifo{P1}{P2}{m}
		\end{tikzcirc}\hphantom{xxxx}%
	}
	\hfil
	\subfloatv[\HEIGHT]{{}\protect\reoname{Merg2}{}{}{p_1 , p_2 ; p_3}}{%
		\hphantom{xxxx}\begin{tikzcirc}
			\port{P1}{0,0}{above}{p_1}
			\port{P2}{$(P1)+(165:\dist cm)$}{above}{p_2}
			\port{P3}{$(P1)+(-165:\dist cm)$}{above}{p_3}
			\merger{P2}{P3}{P1}
		\end{tikzcirc}\hphantom{xxxx}%
	}
	\hfil
	\subfloatv[\HEIGHT]{{}\protect\reoname{Repl2}{}{}{p_1 ; p_2 , p_3}}{%
		\hphantom{xxxx}\begin{tikzcirc}
			\port{P1}{0,0}{above}{p_1}
			\port{P2}{$(P1)+(15:\dist cm)$}{above}{p_2}
			\port{P3}{$(P1)+(-15:\dist cm)$}{above}{p_3}
			\replicator{P1}{P2}{P3}
		\end{tikzcirc}\hphantom{xxxx}%
	}
	\hfil
	\subfloatv[\HEIGHT]{{}\protect\reoname{BinOp}{f}{}{p_1 , p_2 ; p_3}}{%
		\hphantom{xxxxx}\begin{tikzcirc}
			\port{P1}{0,0}{above}{p_1}
			\port{P2}{$(P1)+(165:\dist cm)$}{above}{p_2}
			\port{P3}{$(P1)+(-165:\dist cm)$}{above}{p_3}
			\binop{P2}{P3}{P1}{f}
		\end{tikzcirc}\hphantom{xxxxx}%
	}
	\hfil
	
	\caption{Digraphs for the primitives in Figure~\ref{fig:autom+core}}
	\label{fig:reo+core}
\end{figure}

\begin{figure}[t]
	\captionsetup[subfloat]{labelformat=empty}
	\centering
	\subfloatvspace
	\renewcommand{\reosize}{\scriptsize}
	
	\renewcommand{\HEIGHT}{20pt}%
	\hfil%
	\subfloatvh[\HEIGHT][.5\linewidth]{{}\protect\reoname{Sync$_2$}{}{}{\reo{In} ; \reo{Out}}}{%
		\renewcommand{\reosize}{\tiny}%
		\begin{tikzcirc}
			\port{In}{0,0}{above}{\reo{In}}
			\port[intern]{P}{right of=In}{above}{\reo{P}}
			\port{Out}{right of=P}{above}{\reo{Out}}
			\sync{In}{P}
			\sync{P}{Out}
		\end{tikzcirc}
	}%
	\hfil%
	\subfloatvh[\HEIGHT][.5\linewidth]{{}\protect\reoname{Fifo$_2$}{}{;\reo{x}_1,\reo{x}_2}{\reo{In} ; \reo{Out}}}{%
		\renewcommand{\reosize}{\tiny}%
		\hphantom{xx}\begin{tikzcirc}
			\port{In}{0,0}{above}{\reo{In}}
			\port[intern]{P}{right of=In}{above}{\reo{P}}
			\port{Out}{right of=P}{above}{\reo{Out}}
			\fifo{In}{P}{\reo{x}_1}
			\fifo{P}{Out}{\reo{x}_2}
		\end{tikzcirc}\hphantom{xx}%
	}%
	\hfil%
	
	\renewcommand{\HEIGHT}{137pt}%
	\hfil%
	\hbox{\vbox to \HEIGHT{%
		\renewcommand{\HEIGHT}{43pt}%
		\renewcommand{\WIDTH}{.5\linewidth}%
		\vfil%
		\hbox{\subfloatvh[\HEIGHT][\WIDTH]{{}\protect\reoname{LateAsyncMerg$_2$}{}{;\reo{x}}{\reo{In}_1,\reo{In}_2 ; \reo{Out}}}{%
			\renewcommand{\reosize}{\tiny}%
			\begin{tikzcirc}
				\port[intern]{P}{0,0}{above}{\reo{P}}
				\port{In1}{$(P)+(165:\dist cm)$}{above}{\reo{In}_1}
				\port{In2}{$(P)+(-165:\dist cm)$}{above}{\reo{In}_2}
				\port{Out}{right of=P}{above}{\reo{Out}}
				\merger{In1}{In2}{P}
				\fifo{P}{Out}{\reo{x}}
			\end{tikzcirc}%
		}}%
		\vfil%
		\hbox{\subfloatvh[\HEIGHT][\WIDTH]{{}\protect\reoname{EarlyAsyncMerg$_2$}{}{;\reo{x}_1,\reo{x}_2}{\reo{In}_1,\reo{In}_2 ; \reo{Out}}}{%
			\renewcommand{\reosize}{\tiny}%
			\begin{tikzcirc}
				\port{Out}{0,0}{above}{\reo{Out}}
				\port[intern]{P1}{$(Out)+(165:\dist cm)$}{above}{\reo{P}_1}
				\port[intern]{P2}{$(Out)+(-165:\dist cm)$}{above}{\reo{P}_2}
				\port{In1}{left of=P1}{above}{\reo{In}_1}
				\port{In2}{left of=P2}{above}{\reo{In}_2}
				\fifo{In1}{P1}{\reo{x}_1}
				\fifo{In2}{P2}{\reo{x}_2}
				\merger{P1}{P2}{Out}
			\end{tikzcirc}%
		}}%
		\vfil%
	}}%
	\hfil%
	\hbox{\vbox to \HEIGHT{%
		\vfil%
		\hbox{\subfloath[.5\linewidth]{{}\protect\reoname{Rout$_2$}{}{}{\reo{In} ; \reo{Out}_1 , \reo{Out}_2}}{%
			\renewcommand{\reosize}{\tiny}%
			\hfil%
			\begin{tikzcirc}
				\port{In}{0,0}{above}{\reo{In}}
				\port[intern]{P1}{$(In)+(30:1*\dist cm)$}{above}{\reo{P}_1}
				\port[intern]{P2}{right of=In, node distance=1*\dist cm}{above}{\reo{P}_2}
				\port[intern]{P3}{$(In)+(-30:1*\dist cm)$}{above}{\reo{P}_3}
				
				\port[intern]{P5}{right of=P2}{above}{\reo{P}_5}
				\port[intern]{P7}{$(P5)+(15:1*\dist cm)$}{above}{\reo{P}_7}
				\port[intern]{P8}{$(P5)+(-15:1*\dist cm)$}{above}{\reo{P}_8}
				
				\port[intern]{P4}{$(P7)+(165:\dist cm)$}{above}{\reo{P}_4}
				\port{Out1}{$(P4)+(15:\dist cm)$}{above}{\reo{Out}_1}
				\port[intern]{P6}{$(P8)+(-165:\dist cm)$}{above}{\reo{P}_6}
				\port{Out2}{$(P6)+(-15:\dist cm)$}{above}{\reo{Out}_2}
				
				\replicator{In}{P1}{P3}
				\sync{In}{P2}
				\syncdrain{P2}{P5}
				\merger{P7}{P8}{P5}
				\replicator{P4}{P7}{Out1}
				\replicator{P6}{P8}{Out2}
				\lossysync{P1}{P4}
				\lossysync{P3}{P6}
			\end{tikzcirc}%
			\hfil%
		}}%
		\vfil%
	}}%
	\hfil%
	
	\subfloat[{}\protect\reoname{OddFib$_2$}{}{;\reo{x}_1,\reo{x}_2,\reo{y}_1,\reo{y}_2}{\reo{In} ; \reo{Out}_1 , \reo{Out}_2}]{%
		\renewcommand{\reosize}{\tiny}%
		\begin{tikzcirc}
			\port[intern]{E}{0,0}{right}{\reo{E}}
			\port[intern]{B}{above of=E}{above}{\reo{B}}
			\port[intern]{D}{above left of=E}{left}{\reo{D}}
			\binop{B}{D}{E}{\code{add}}
			
				\port[intern]{F}{below of=E}{below}{\reo{F}}
				\port[intern]{G}{below right of=E}{above}{\reo{G}}
				\replicator{E}{F}{G}
				
					\port[intern]{P1}{left of=F}{below}{\reo{P}_1}
					\fifo{F}{P1}{\reo{y}_1}
					
						\port[intern]{P2}{left of=P1}{below}{\reo{P}_2}
						\fifod{P1}{P2}{\reo{y}_2}{\code{1}}
						
							\port[intern]{C}{above right of=P2}{right}{\reo{C}}
							\port[intern]{P3}{above of=P2}{left}{\reo{P}_3}
							\replicator{P2}{C}{P3}
							\sync{C}{D}
							
								\port[intern]{P4}{above left of=P3}{below}{\reo{P}_4}
								\port[intern]{P5}{above of=P3}{above}{\reo{P}_5}
								\replicator{P3}{P4}{P5}
								
									\port[intern]{A}{right of=P5}{above}{\reo{A}}
									\fifod{P5}{A}{\reo{x}_1}{\code{0}}
									\fifo{A}{B}{\reo{x}_2}
									
									\port{In}{left of=P4}{below}{\reo{In}}
									\syncdrain{P4}{In}
					
					\port[intern]{H}{right of=G}{above}{\reo{H}}
					\filter{G}{H}{\code{Odd}}
					
						\port{Out1}{$(H)+(15:\dist cm)$}{above}{\reo{Out}_1}
						\port{Out2}{$(H)+(-15:\dist cm)$}{above}{\reo{Out}_2}
						\replicator{H}{Out1}{Out2}
		\end{tikzcirc}
	}
	
	\caption{Digraphs for six example composites}
	\label{fig:reo+oddfibonacci+2}
\end{figure}

\noindent Instead of writing explicit~$\otimes/\ominus$-expressions to construct \CAs, in practice, we often draw them in a graphical, more intuitive syntax, based on the coordination language Reo~\cite{Arb04,Arb11}.\footnote{%
	Other syntaxes for \CAs beside Reo exist.
	For instance, we know how to translate \textsc{Uml} sequence/activity diagrams and \textsc{Bpmn} to \CAs~\cite{AKM08,CKA10,MAB11}.
	Connector algebras of Bliudze and Sifakis~\cite{BS10} also have a straightforward interpretation in terms of \CAs, so offering an interesting alternative syntax~\cite{DJAB15}.
}
Essentially, in this syntax, we draw a (hyper)digraph, where every vertex denotes a port, and where every (hyper)arc denotes a \CA consisting of the ports denoted by its connected vertices.
By convention, every vertex has degree~$1$ (for input and output ports) or~$2$ (for internal ports).
The~$\otimes/\ominus$-expression denoted by a digraph, then, is the join of (the denotations of) its arcs, and the hide of (the denotations of) its vertices of degree~$2$.
Intuitively, every transition in the (evaluated)~$\otimes/\ominus$-expression for a digraph corresponds to an atomic flow of data along the arcs in that digraph.
Figure~\ref{fig:reo+core} shows digraphs for the primitives in Figure~\ref{fig:autom+core}; Figure~\ref{fig:reo+oddfibonacci+2} shows digraphs for example composites.

In Figure~\ref{fig:reo+oddfibonacci+2}, \reo{Sync}$_2$ models the same coordinator as a single \reo{Sync}: it enforces a standard synchronous channel protocol between a producer and a consumer.
\reo{Fifo}$_2$ models a coordinator between a producer and a consumer that enforces a standard (order-preserving) asynchronous channel protocol with a buffer of capacity~$2$.
\reo{LateAsyncMerg}$_2$ is (a behaviorally congruent \CA to) the \CA in Figure~\ref{fig:autom+lateasyncmerger2}.
\reo{EarlyAsyncMerg}$_2$ models a coordinator between two producers and one consumer, as \reo{LateAsyncMerg}$_2$.
The difference between the two is that with \reo{EarlyAsyncMerg}$_2$, every producer has its own buffer, which results in significantly different behavior (as producers no longer need to wait for each other before their \code{put}s can complete).
\reo{Rout}$_2$ models a coordinator between one producer and two consumers that enforces a symmetric protocol to \reo{Merg2}: infinitely often, it atomically \SCOPE{accepts a datum on its input port, then offers it on one of its output ports}.
Finally, \reo{OddFib}$_2$ models a coordinator between two producers and one consumer.
Whenever the~$i$-th \code{put} by the producer completes, one of two things happens.
If the~$i$-th Fibonacci number is even, the datum \code{put} by the producer is lost, and no interaction occurs between the producer and the two consumers.
If the~$i$-th Fibonacci number is odd, in contrast, a \code{get} by each of the two consumers \emph{must} complete at the same time (i.e., atomically, i.e., synchronously).
In this case, specifically, the datum \code{put} by the producer is lost, while the consumers \code{get} the~$i$-th Fibonacci number.
This coordinator, thus, enforces synchronous, unreliable (in the sense just described) communication from a producer to two consumers.

\edit{The primitives in Figure~\ref{fig:reo+core} were introduced by Arbab~\cite{Arb04}, except \reo{BinOp}, which was introduced by Jongmans~\cite{Jon16} (\reo{BinOp} is, however, a generalization of primitive \reo{Join}, which was introduced by Kokash and Arbab~\cite{KA09}).
\reo{LateAsyncMerg} and \reo{EarlyAsyncMerg} in Figure~\ref{fig:reo+oddfibonacci+2} are probably folklore; these two names were first used by Jongmans~\cite{Jon16}.
\reo{OddFib} is based on Arbab's \reo{Fibonacci}~\cite{Arb05}.
\reo{Rout} was introduced by Arbab~\cite{Arb05}.}

%
\section{Optimization I: Eliminate (Instead of Hide)}
\label{sect:elim}
\subsection*{Motivating Example}

To illustrate the need for our first technique to optimize the performance of checking data constraints, presented in this section, we start with a motivating example.
Recall the \reo{Sync} primitive in Figure~\ref{fig:autom+core}.
\reo{Sync} has a special property: it acts as a kind of algebraic identity of join and hide, in the following sense.
Let~${\bfa} [p' / p]$ denote \CA~$\bfa$ with port~$p'$ substituted for every occurrence of port~$p$.
Let~$\bfa$ range over the set of all \CAs that \edit{(i)} have an input port~$p_2$ \edit{\emph{and} (ii) in which port~$p_1$ does not occur}.
Then:
$$
(\reoname{Sync}{}{}{p_1 ; p_2} \multx \bfa) \subtrx p_2 \congr {\bfa} [p_1 / p_2]
$$
In words,~$(\reoname{Sync}{}{}{p_1 ; p_2} \multx \bfa) \subtrx p_2$ and~$\bfa$ are behaviorally congruent modulo substitution of~$p_1$ for~$p_2$.
Generally, we can ``prefix'' (i.e., join on its input ports) or ``suffix'' (i.e., join on its output ports) any number of \reo{Sync}s to a \CA without affecting---in the sense just described---that \CA's behavior.
Given this property, it seems not unreasonable to assume that compiler-generated code for a single \reo{Sync} has the same performance as a chain of~$64$ \reo{Sync}s.
Slightly more formally, if~$\sim$ means ``has the same performance'', one may expect:
$$
\reoname{Sync}{}{}{p_1 ; p_{65}} \sim (\reoname{Sync}{}{}{p_1 ; p_2} \multx \cdots \multx \reoname{Sync}{}{}{p_{64} ; p_{65}}) \subtrx p_2 \subtrx \cdots \subtrx p_{64}
$$
Our \compilergenerated code, however, violates this equation: a single \reo{Sync} fires~$27$ million transitions in four minutes, whereas the chain of 64 \reo{Sync}s fires only nine million transitions.

To understand this phenomenon, we first present the definition of hide~\cite{BSAR06,Jon16}:

\begin{defi}
	[hide]
	\label{def:subtrx}
	$\subtrx : \automuniv \times \puniv \rightarrow \automuniv$ denotes the function defined by the following equation:
	$$
	\tpl{Q \, \tpl{P^\textall \, P^\textin \, P^\textout} \, M \, \longrightarrow \, q^0} \subtrx p =  \tpl{Q \, \tpl{P^\textall \setminus \set{p} \, P^\textin \setminus \set{p} \, P^\textout \setminus \set{p}} \, M \, \longrightarrow_\subtrx \, q^0}
	$$
	where~$\longrightarrow_\subtrx$ denotes the smallest relation induced by the following rule:
	\begin{equation}
		\dfrac{%
			q \xrightarrow{P,\phi} q'
		}{%
			q \xrightarrow{P\setminus\set{p},\exists p.\phi}_\subtrx q'
		}
	\end{equation}
\end{defi}

\noindent
In words, hide removes a port both from sets~$P^\textall , P^\textin , P^\textout$ and from every transition.
\edit{(Because $P^\textin , P^\textout \subseteq P^\textall$ by Definition~\ref{def:automuniv}, we need to remove~$p$ not only from~$P^\textin$ and~$P^\textout$ but also from~$P^\textall$.)}
But whereas hide removes ports from synchronization constraints \emph{syntactically}---ef\-fec\-tive\-ly making those constraints smaller---it removes ports from data constraints only \emph{semantically}.
Indeed,~$\subtrx$ does not reduce the size of data constraints (in terms of the number of data variables, data literals, and existential quantifications) but, in fact and in contrast, makes data constraints larger by enveloping them in existential quantifications: the transition in the single \reo{Sync} has just~$p_1 \termeq p_{65}$ as its data constraint, whereas the corresponding transition in the chain of~$64$ \reo{Sync}s has~$\exists p_{64} . \cdots \exists p_2 . (p_1 \termeq p_2 \wedge \cdots \wedge p_{64} \termeq p_{65})$.
Clearly, \editt{although the two data constraint expressions are semantically (logically) equivalent,} checking the latter data constraint \editt{expression} requires more resources than the former.

Below, we develop a variant of hide, called \emph{eliminate}, that, when applied~$63$ times to the chain of~$64$ \reo{Sync}s, yields the same data constraint as the one in the single \reo{Sync}.
\edit{The key idea is to mechanically simplify data \editt{constraint expressions} using \editt{the} equivalence~$\exists p . (p \termeq t \wedge \phi) \dcequiv \phi [t / p]$, if~$p \notin \freefun(t)$, \editt{whenever} this equivalence becomes applicable after hiding.
In the previous example, for instance, we can use this equivalence to simplify~$\exists p_{64} . \cdots \exists p_3 . \exists p_2 . (p_1 \termeq p_2 \wedge p_2 \termeq p_3 \wedge \cdots \wedge p_{64} \termeq p_{65})$ to~$\exists p_{64} . \cdots \exists p_3 . (p_1 \termeq p_3 \wedge \cdots \wedge p_{64} \termeq p_{65})$.
We can subsequently repeat this process until we indeed arrive at \editt{the expression}~$p_1 \termeq p_{65}$, as desired.}

\subsection*{Eliminate}

First, we need to introduce the concept of \emph{determinants} of free data variables in data constraints.
For a data constraint~$\varphi$ and one of its free data variables~$x \in \freefun(\varphi)$, the set of determinants of~$x$ consists of those terms that \emph{precisely} determine the datum~$\sigma(x)$ assigned to~$x$ in any data assignment~$\sigma$ that satisfies~$\varphi$ (i.e.,~$\sigma \dcmodels \varphi$).
``Precisely'' here means that a determinant neither overspecifies nor underspecifies~$\sigma(x)$.
Thus, if a set of determinants contains multiple data terms, each of those data terms evaluates to the same datum under~$\sigma$.
Determinants furthermore determine~$\sigma(x)$ independent of~$x$ itself: no determinant of~$x$ has~$x$ among its free data variables (i.e., determinants have no recursion).

\begin{defi}
	[determinants]
	\label{def:determfun}
	$\determfun : \xuniv \times \dcuniv \rightarrow \pset{\termuniv}$ denotes the function defined by the following equations:
	$$
	\begin{array}{@{} l @{\ } c @{\ } l @{}}
		\determfun_x(\top) \, \determfun_x(\bot)			& =	& \emptyset
	\\	\UP
	\\	\determfun_x(t_1 \termeq t_2)						& =	& \left \{ \begin{array}{@{} l @{} l @{}}
																	\set{t_2}	& \IF \SCOPE{t_1 = x \AND x \notin \variablfun(t_2)}
																\\	\set{t_1}	& \IF \SCOPE{t_2 = x \AND x \notin \variablfun(t_1)}
																\\	\emptyset	& \OTHERWISE
																\end{array} \right .
	\\	\UP
	\\	\determfun_x(R(t_1 \, \ldots \, t_k))				& = & \emptyset
	\\	\determfun_x(\neg a)								& = & \emptyset
	\\	\determfun_x(\ell_1 \wedge \cdots \wedge \ell_k)	& =	& \determfun_x(\ell_1) \cup \cdots \cup \determfun_x(\ell_k)
	\\	\UP
	\\	\determfun_x(\exists x' . \varphi')					& =	& \left \{ \begin{array}{@{} l @{} l @{}}
																	\determfun_x(\varphi)	& \IF x \neq x'
																\\	\emptyset				& \OTHERWISE
																\end{array} \right .
	\end{array}
	$$
\end{defi}

\noindent
For instance, consider the following data constraint:\label{page:phieg}
$$
\phieg = \pre{\reo{x}_2} \termeq \reo{B} \wedge \reo{C} \termeq \reo{D} \wedge \code{add}(\reo{B} , \reo{D}) \termeq \reo{E} \wedge \reo{E} \termeq \reo{F} \wedge \reo{E} \termeq \reo{G} \wedge \neg \code{Odd}(\reo{G})
$$
(This data constraint appears in the~$\otimes/\ominus$-expression denoted by the digraph for \reo{OddFib} in Figure~\ref{fig:reo+oddfibonacci+2}.)
The free data variables in~$\phieg$ have the following determinants:
$$
\begin{array}{@{} l @{\ } c @{\ } l @{}}
	\determfun_\pre{\reo{x}}(\phieg)							& =	& \set{\reo{B}}
\\	\determfun_\reo{B}(\phieg)									& =	& \set{\pre{\reo{x}}}
\\	\determfun_\reo{C}(\phieg)									& =	& \set{\reo{D}}
\\	\determfun_\reo{D}(\phieg)									& =	& \set{\reo{C}}
\end{array} \qquad \begin{array}{@{} l @{\ } c @{\ } l @{}}
	\determfun_\reo{E}(\phieg)									& =	& \set{\code{add}(\reo{B} \, \reo{D}) \, \edit{\reo{F}} \, \edit{\reo{G}}}
\\	\determfun_\reo{F}(\phieg)									& =	& \set{\reo{E}}
\\	\determfun_\reo{G}(\phieg)									& =	& \set{\reo{E}}
\end{array}
$$\medskip

\noindent Next, let~$\bfa$ denote a \CA, and let~$\varphi$ denote one of its data constraints.
Suppose that we hide~$x$ from~$\bfa$ with~$\subtrx$.
By Definition~\ref{def:subtrx} of~$\subtrx$, the transition(s) of~$\bfa$ previously labeled by~$\varphi$ are now labeled with~$\exists x . \varphi$.
However, if~$x$ has determinants, instead of enveloping~$\varphi$ in an existential quantification as~$\subtrx$ does, we can alternatively perform a syntactic substitution of one of those determinants for~$x$.
We formalize such a substitution as follows.

\begin{defi}
	[syntactic existential quantification]
	\label{def:existsfun}
	$\existsfun : \xuniv \times \dcuniv \rightarrow \dcuniv$ denotes the function defined by the following equation:
	$$
	\LINES[l]{%
		\existsfun_x(\varphi) = \left \{ \begin{array}{@{} l @{} l @{}}
			\varphi \subst{t}{x}	& \IF \SCOPE{\determfun_x(\varphi) \neq \emptyset \AND t = \min{\determfun_x(\varphi)}}
		\\	\exists x . \varphi		& \OTHERWISE
		\end{array} \right .
	}
	$$
\end{defi}

\noindent
In this definition, function~$\min{\cdot}$ takes the least element in~$\determfun_x(\varphi)$, under the global order on data terms~$\termorder$, to ensure that~$\existsfun$ always produces the same output under the same input.
The following equations exemplify the (nested) application of~$\existsfun$ on~$\phieg$.
$$
\begin{array}{@{} l @{\ } l @{}}
		& \existsfun_\reo{G}(\existsfun_\reo{E}(\existsfun_\reo{D}(\existsfun_\reo{B}(\phieg))))
\\	\UP
\\	=	& \existsfun_\reo{G}(\existsfun_\reo{E}(\existsfun_\reo{D}(\existsfun_\reo{B}(
\\		& \quad \pre{\reo{x}} \termeq \reo{B} \wedge \reo{C} \termeq \reo{D} \wedge \code{add}(\reo{B} \, \reo{D}) \termeq \reo{E} \wedge \reo{E} \termeq \reo{F} \wedge \reo{E} \termeq \reo{G} \wedge \neg \code{Odd}(\reo{G})))))
\\ \UP
\\	=	& \existsfun_\reo{G}(\existsfun_\reo{E}(\existsfun_\reo{D}(
\\		& \quad \lgray \pre{\reo{x}} \termeq \pre{\reo{x}} \wedge \black \reo{C} \termeq \reo{D} \wedge \code{add}(\pre{\reo{x}} \, \reo{D}) \termeq \reo{E} \wedge \reo{E} \termeq \reo{F} \wedge \reo{E} \termeq \reo{G} \wedge \neg \code{Odd}(\reo{G}))))
\\ \UP
\\	=	& \existsfun_\reo{G}(\existsfun_\reo{E}(
\\		& \quad \lgray \pre{\reo{x}} \termeq \pre{\reo{x}} \wedge \reo{C} \termeq \reo{C} \wedge \black \code{add}(\pre{\reo{x}} \, \reo{C}) \termeq \reo{E} \wedge \reo{E} \termeq \reo{F} \wedge \reo{E} \termeq \reo{G} \wedge \neg \code{Odd}(\reo{G})))
\\	\UP
\\	=	& \existsfun_\reo{G}(
\\		& \quad \lgray \pre{\reo{x}} \termeq \pre{\reo{x}} \wedge \reo{C} \termeq \reo{C} \wedge \black \code{add}(\pre{\reo{x}} \, \reo{C}) \termeq \reo{F} \wedge \lgray \reo{F} \termeq \reo{F} \wedge \black \reo{F} \termeq \reo{G} \wedge \neg \code{Odd}(\reo{G}))
\\	\UP
\\	=	& \quad \lgray \pre{\reo{x}} \termeq \pre{\reo{x}} \wedge \reo{C} \termeq \reo{C} \wedge \black \code{add}(\pre{\reo{x}} \, \reo{C}) \termeq \reo{F} \wedge \lgray \reo{F} \termeq \reo{F} \wedge \reo{F} \termeq \reo{F} \black \wedge \neg \code{Odd}(\reo{F})
\end{array}
$$
We define eleminate in terms of~$\existsfun$.

\begin{defi}
	[eliminate]
	\label{def:subtry}
	$\subtry : \automuniv \times \puniv \rightarrow \automuniv$ denotes the function defined by the following equation:
	$$
	\tpl{Q \, \tpl{P^\textall \, P^\textin \, P^\textout} \, M \, \longrightarrow \, q^0} \subtry p = \tpl{Q \, \tpl{P^\textall \setminus \set{p} \, P^\textin \setminus \set{p} \, P^\textout \setminus \set{p}} \, M \, \longrightarrow_\subtry \, q^0}
	$$
	where~$\longrightarrow_\subtry$ denotes the smallest relation induced by the following rule:
	\begin{equation}
		\dfracx[l]{%
			q \xrightarrow{P,\varphi} q'
		}{%
			q \xrightarrow{P\setminus\set{p},\existsfun_p(\varphi)}_\subtry q'
		}
	\end{equation}
\end{defi}\medskip

\noindent
In the previous definition, we use~$\existsfun$ to remove ports from data constraints.
Although Definition~\ref{def:existsfun} of~$\existsfun$ also allows for removing data variables for memory cells, we do not pursue such elimination in this paper.

%
\subsection*{Correctness and Effectiveness}

We conclude this section by establishing the correctness and effectiveness of eliminate.
We consider eliminate correct if it yields a \editt{\CA} behaviorally congruent to the \CA that hide \editt{yields}.
Before formulating this as a theorem, the following lemma first states the equivalence of existential quantification and~$\existsfun$.

\begin{lem}
	\label{lemma:dcequiv+exists+existsfun}
	$\exists x . \varphi \dcequiv \existsfun_x(\varphi)$
\end{lem}

From Proposition~\ref{prop:001} and Lemma~\ref{lemma:dcequiv+exists+existsfun}, we conclude the following correctness theorem.

\begin{thm}
	\label{thm:congr+subtrx+subtry}
	$\bfa \subtrx p \congr \bfa \subtry p$
\end{thm}

We consider eliminate effective if, after eliminating a port~$p$ from a \CA~$\bfa$, that port no longer occurs in any of that \CA's data \editt{constraint expressions}.
Generally, however, such unconditional effectiveness does not hold true: if~$\bfa$ has a data constraint~$\varphi$ in which~$p$ occurs, but~$p$ has no determinants in~$\varphi$, eliminate has nothing to replace~$p$ with.
In that case,~$\existsfun_p(\varphi) = \exists p . (\varphi)$, and consequently, eliminate does not have its intended \editt{(simplifying)} effect.
Eliminate \emph{does} satisfy a weaker---but useful---form of effectiveness, though.
To formulate this as a theorem, we first define a function that computes \emph{\everdetermined ports}.
We call a port~$p$ \everdetermined in a \CA~$\bfa$ iff both~$p$ occurs in~$\bfa$ and every data constraint in~$\bfa$ has a determinant for~$p$.

\begin{defi}
	[\everdetermined ports]
	\label{def:edpfun}
	$\edpfun : \automuniv \rightarrow \pset{\puniv}$ denotes the function defined by the following equation:
	$$
	\edpfun(\bfa) = \setb{p}{\SCOPE{\SCOPEX[l]{%
		p \in \variablfun(\varphi)
	\\	\LAND \varphi \in \dcfun(\bfa)
	} \IMPLIES \determfun_p(\varphi) \neq \emptyset} \FORALL \varphi}
	$$
\end{defi}

\noindent
For instance,~$p_1$,~$p_2$, and~$p_3$ all qualify as \everdetermined in \reo{Merg2} in Figure~\ref{fig:autom+core}.
To understand the \everdeterminedness of~$p_1$, observe that~$p_1$ occurs in the data constraint on the top transition in \reo{Merg2} and that~$p_1$ has a determinant in that data constraint (namely~$p_3$); because~$p_1$ does not occur in the data constraint on the bottom transition in \reo{Merg2},~$p_1$ indeed qualifies as \everdetermined.
A similar explanation applies to~$p_2$.
To understand the \everdeterminedness of~$p_3$, observe that~$p_3$ occurs in the data constraint on both transitions in \reo{Merg2} and that~$p_3$ has a determinant in both these data constraints (namely~$p_1$ and~$p_2$).
Consequently, also~$p_3$ qualifies as \everdetermined.
In contrast,~$\edit{p_1}$ in members of \reo{Filter} in Figure~\ref{fig:autom+core} does \emph{not} qualify as \everdetermined, because~$\edit{p_1}$ occurs in the data constraint on the top transition in \reo{Filter} but does not have a single determinant in that data constraint.

The following theorem states the effectiveness of eliminate, conditional on \everdeterminedness: after eliminating an \everdetermined port from a \CA, that port no long occurs in any of that \CA's data constraints.

\begin{thm}
	\label{thm:subtry+effect}
	$p \in \edpfun(\bfa) \IMPLIES p \notin \setb{x}{\varphi \in \dcfun(\bfa \subtry p) \AND x \in \variablfun(\varphi)}$
\end{thm}

``Effectiveness'' refers to a rather theoretical property; it says nothing yet about the impact of applying~$\subtry$ in practice.
In Section~\ref{sect:exper}, we study this impact through a number of experiments; in this section, we only revisit our \edit{motivating} example.
By using~$\subtry$ instead of~$\subtrx$, and after removing~$t \termeq t$ literals (each of which trivially equates to~$\top$), we get exactly the same data constraint in the chain of~$64$ \reo{Sync}s as in the single \reo{Sync}.
Consequently, the \compilergenerated code for the chain of~$64$
\reo{Sync}s has the same performance as \compilergenerated code for
the single \reo{Sync} (which corresponds to a~$3\times$ speedup
relative to unoptimized code generated with hide instead of
eliminate).

%
\section{Optimization II: Commandify (Instead of Seek)}
\label{sect:comm}
\subsection*{Data Commands}

In the previous section, we presented a first technique to optimize the performance of checking data constraints.
In this section, we present a second technique to further optimize the performance of such checks and, in particular, the expensive constraint solver calls involved.
Essentially, this new technique comprises the generation of a little, dedicated constraint solver for every data constraint at compile-time.
At run-time, then, instead of calling a general-purpose constraint solver to check a data constraint, the \compilergenerated coordinator thread for a \CA calls a more efficient constraint solver generated specifically for that data constraint.
First, in this subsection, we describe a basic sequential language (syntax, semantics, proof system) in which to express such dedicated constraint solvers; in the next subsections, we present the process of their generation.

General-purpose techniques for constraint solving---an \textsc{np}-com\-plete problem for finite domains---inflict not only \editt{a solving} overhead proportional to the size of a data constraint but also a constant overhead for preparing, making, and processing the result of every call \editt{to a full-fledged solver}.
Although we generally cannot escape using such techniques for checking arbitrary data constraints, a better alternative exists for many data constraints in practice.
The crucial observation is that the data constraints in \emph{all} \CAs that we know of in the literature really constitute declarative specifications of a relatively straightforward imperative program.
What we need to do, then, is develop a technique for statically translating such a data constraint~$\varphi$, off-line at compile-time, into a \editt{small} imperative program that computes a data assignment~$\sigma$ such that~$\sigma \dcmodels \varphi$, without \editt{resorting to} general-purpose constraint solving.
We call such a \editt{small} program a \emph{data command} and the translation from data constraints to data commands \emph{commandification}.
Essentially, we formalize and automate what programmers do when they write an imperative implementation of a declarative specification expressed as a data constraint.
After presenting our technique, we make the class of data constraints currently supported by commandification precise.

\begin{defi}
	[data commands]
	\label{def:communiv}
	A data command is an object generated by the following grammar:
	$$
	\begin{array}{@{} l @{\ } c @{\ } l @{\qquad} l @{}}
		\pi	& ::=	& \commskip \mid x \commasgn t \mid \commifthen{\varphi}{\pi} \mid \pi \commseq \pi \mid \varepsilon	& \text{(data commands)}
	\end{array}
	$$
	$\communiv$ denotes the set of all data commands.
\end{defi}

In the previous definition,~$\varepsilon$ denotes the empty data command,~$x \commasgn t$ denotes an \emph{assignment}, and~$\commifthen{\varphi}{\pi}$ denotes a \emph{failure statement}.\footnote{%
	\edit{The term ``failure statement'' may be confusing.
	As shortly formalized in Definition~\ref{def:transsyst}, it refers to a special conditional statement that fails in case the first alternative cannot be selected.
	By calling such statements ``failure statements'', we follow Apt et al.~\cite{AdBO09}, which strongly influenced this section.}
}
Henceforth, we often write ``value of~$x$'' instead of ``the datum assigned to~$x$''.

We define an operational semantics for data commands based on an operational semantics for a sequential language by Apt et al.~\cite{AdBO09}.
\edit{As} data commands \edit{are supposed to} solve data constraints, we model the \emph{data state} that a data command executes in with either a function from data variables to data---a data assignment---or the distinguished object~$\fail$, which models abnormal termination.
A \emph{data configuration}, then, consists of a data command and a data state to execute that data command in.

\begin{defi}
	[abnormal termination]
	\label{def:fail}
	$\fail$ is an unstructured object such that~$\fail \notin \assignmuniv$.
\end{defi}

\begin{defi}
	[data configurations]
	A data configuration is a pair~$\tpl{\pi \, \varsigma}$ where:
	\begin{itemize}
		\item
		$\pi \in \communiv$ \hfill (data command)
		
		\item
		$\varsigma \in \assignmuniv \cup \set{\fail}$ \hfill (data state)
	\end{itemize}
	$\confuniv$ denotes the set of all data configurations.
\end{defi}

A \emph{transition system} on configurations formalizes their evolution in time.

\begin{figure}
	\centering
	
	\renewcommand{\WIDTH}{.5\linewidth}
	\begin{framed}
		\begin{tabular}{@{} c @{} c @{}}
			\begin{minipage}{\WIDTH}
				\noindent%
				\begin{equation}
					\qquad \dfrac{}{\tpl{\commskip \, \sigma} \transsyst \tpl{\varepsilon \, \sigma}} \qquad
				\end{equation}
			\end{minipage}
		&	\begin{minipage}{\WIDTH}
				\noindent%
				\begin{equation}
					\dfrac{}{\tpl{x \commasgn t \, \sigma} \transsyst \tpl{\varepsilon \, \sigma [x \mapsto \evalfun_\sigma(t)]}}
				\end{equation}
			\end{minipage}
		\\
		\\	\begin{minipage}{\WIDTH}
				\noindent%
				\begin{equation} 
					\quad \dfrac{%
						\sigma \dcmodels \varphi
					}{%
						\tpl{\commifthen{\varphi}{\pi} \, \sigma} \transsyst \tpl{\pi \, \sigma}
					} \quad
				\end{equation}
			\end{minipage}
		&	\begin{minipage}{\WIDTH}
				\noindent%
				\begin{equation} 
					\quad \dfrac{%
						\sigma \not \dcmodels \varphi
					}{%
						\tpl{\commifthen{\varphi}{\pi} \, \sigma} \transsyst \tpl{\varepsilon \, \fail}
					} \quad
				\end{equation}
			\end{minipage}
		\\
		\\	\begin{minipage}{\WIDTH}
				\noindent%
				\begin{equation} 
					\dfrac{%
						\tpl{\pi \, \sigma} \transsyst \tpl{\pi' \, \sigma'} \AND \pi' \neq \varepsilon
					}{%
						\tpl{\pi \commseq \pi'' \, \sigma} \transsyst \tpl{\pi' \commseq \pi'' \, \sigma'}
					}
				\end{equation}
			\end{minipage}
		&	\begin{minipage}{\WIDTH}
				\noindent%
				\begin{equation}
					\quad \dfrac{%
						\tpl{\pi \, \sigma} \transsyst \tpl{\varepsilon \, \sigma'}
					}{%
						\tpl{\pi \commseq \pi'' \, \sigma} \transsyst \tpl{\pi'' \, \sigma'}
					} \quad
				\end{equation}
			\end{minipage}
		\end{tabular}
	\end{framed}
	
	\caption{Addendum to Definition~\ref{def:transsyst}}
	\label{fig:def+transsyst}
\end{figure}

\begin{defi}
	[transition system on data configurations]
	\label{def:transsyst}
	${\transsyst} \subseteq \confuniv \times \confuniv$ denotes the smallest relation induced by the rules in Figure~\ref{fig:def+transsyst}.
\end{defi}

\noindent
Note that~$\commifthen{\varphi}{\pi}$ indeed denotes a failure statement rather than a \emph{conditional statement}: if the current data state violates the \emph{guard}~$\varphi$, execution abnormally terminates.

Through the transition system in Definition~\ref{def:transsyst}, we associate two different semantics with data commands.
The \emph{partial correctness semantics} of a data command~$\pi$ under a set of \emph{initial} data states~$\Sigma$ consists of all the \emph{final} data states~$\Sigma'$ to which any of those initial states may evolve through execution of~$\pi$.
Notably, this partial correctness semantics ignores abnormal termination.
In contrast, the \emph{total correctness semantics} of~$\pi$ under~$\Sigma$ consists not only of~$\Sigma'$ but, if at least one execution abnormally terminates, also of~$\fail$.

\begin{defi}
	[correctness semantics of data commands]
	\label{def:finalfun}
	\label{def:finalfailfun}
	$\finalfun$, respectively, $\finalfailfun$ denote the functions $\communiv \times \pset{\assignmuniv} \rightarrow \pset{\assignmuniv \cup \set{\fail}}$ defined by the following equations:
	$$
	\begin{array}{@{} l @{\ } c @{\ } l @{}}
		\finalfun(\pi \, \Sigma)		& =	& \setb{\sigma'}{\sigma \in \Sigma \AND \tpl{\pi \, \sigma} \transsyst^* \tpl{\varepsilon \, \sigma'}}
	\\	\finalfailfun(\pi \, \Sigma)	& =	& \finalfun(\pi \, \Sigma) \cup \setb{\fail}{\sigma \in \Sigma \AND \tpl{\pi \, \sigma} \transsyst^* \tpl{\pi' \, \fail}}
	\end{array}
	$$
\end{defi}\medskip

\noindent
Apt et al. showed that all programs from a superset of the set of all data commands execute deterministically~\cite{AdBO09}.
Consequently, also data commands execute deterministically.

\begin{prop}
	[Lemma 3.1 in {\cite[Section 3.2]{AdBO09}}]
	\label{lemma:finalfun+leq1}
	\label{lemma:finalfailfun+eq1}
	\strut
	\begin{itemize}
		\item
		$\card{\finalfun(\pi \, \set{\sigma})} \leq 1$
		
		\item
		$\card{\finalfailfun(\pi \, \set{\sigma})} = 1$
	\end{itemize}
\end{prop}

To prove the correctness of commandification, we use Hoare logic~\cite{Hoa69}, where \emph{triples} of the form~$\tripl{\varphi}{\pi}{\varphi'}$ play a central role.
In such a triple, \emph{precondition}~$\varphi$ characterizes the set of initial data states,~$\pi$ denotes the data command to execute on those states, and \emph{postcondition}~$\varphi'$ characterizes the set of final data states after executing~$\pi$.

\begin{defi}
	[triples]
	$\tripluniv = \dcuniv \times \communiv \times \dcuniv$ denotes the set of all triples, typically denoted by~$\tripl{\varphi}{\pi}{\varphi'}$.
\end{defi}

Let~$\dcchar{\varphi}$ denote the set of data states that satisfy~$\varphi$ (i.e., the data assignments characterized by~$\varphi$).
We interpret triples in two senses: that of partial correctness and that of total correctness.
In the former case, a triple~$\tripl{\varphi}{\pi}{\varphi'}$ holds true iff every final data state to which an initial data state characterized by~$\varphi$ can evolve under~$\pi$ satisfies~$\varphi'$; in the latter case, additionally, execution of~$\pi$ does not abnormally terminate.

\begin{defi}
	[interpretation of triples]
	\label{def:triplmodels}
	\label{def:triplmodelstot}
	${\triplmodels} \, {\triplmodelstot} \subseteq \tripluniv$ denote the smallest relations induced by the following rules:
	\begin{center}
		\begin{tabular}{@{}c@{}c@{}}
			\begin{minipage}{.5\linewidth}
				\noindent%
				\begin{equation}
					\enspace \dfrac{%
						\finalfun(\pi \, \dcchar{\varphi}) \subseteq \dcchar{\varphi'}
					}{%
						\triplmodels \tripl{\varphi}{\pi}{\varphi'}
					} \enspace
				\end{equation}
			\end{minipage}
		&	\begin{minipage}{.5\linewidth}
				\noindent%
				\begin{equation}
					\dfrac{%
						\finalfailfun(\pi \, \dcchar{\varphi}) \subseteq \dcchar{\varphi'}
					}{%
						\triplmodelstot \tripl{\varphi}{\pi}{\varphi'} \vphantom{\triplmodels}
					}
				\end{equation}
			\end{minipage}
		\end{tabular}
	\end{center}
\end{defi}\medskip

\noindent To prove properties of data commands, we use the following sound \emph{proof systems} for partial and total correctness, adopted from Apt et al. with some minor cosmetic changes~\cite{AdBO09}.

\begin{figure}
	\centering
	
	\renewcommand{\WIDTH}{.5\linewidth}
	\begin{framed}
		\begin{tabular}{@{} c @{} c @{}}
			\begin{minipage}{\WIDTH}
				\noindent%
				\begin{equation}
					\label{rule:triplproves+skip}
					\hphantom{xxx} \dfrac{}{\triplproves \tripl{\varphi}{\commskip}{\varphi}} \hphantom{xxx}
				\end{equation}
			\end{minipage}
		&	\begin{minipage}{\WIDTH}
				\noindent%
				\begin{equation}
					\label{rule:triplprovestot+skip}
					\hphantom{xxx} \dfrac{}{\triplprovestot \tripl{\varphi}{\commskip}{\varphi}} \hphantom{xxx}
				\end{equation}
			\end{minipage}
		\\
		\\	\begin{minipage}{\WIDTH}
				\noindent%
				\begin{equation}
					\label{rule:triplproves+asgn}
					\dfrac{}{\triplproves \tripl{\varphi [t / x]}{x \commasgn t}{\varphi}}
				\end{equation}
			\end{minipage}
		&	\begin{minipage}{\WIDTH}
				\noindent%
				\begin{equation}
					\label{rule:triplprovestot+asgn}
					\quad \dfrac{}{\triplprovestot \tripl{\varphi [t / x]}{x \commasgn t}{\varphi}} \quad
				\end{equation}
			\end{minipage}
		\\
		\\	\begin{minipage}{\WIDTH}
				\noindent%
				\begin{equation}
					\label{rule:triplproves+seq}
					\dfracx[r]{%
						\triplproves \tripl{\varphi_1}{\pi_1}{\varphi}
					\\	\LAND \triplproves \tripl{\varphi}{\pi_2}{\varphi_2}
					}{%
						\triplproves \tripl{\varphi_1}{\pi_1 \commseq \pi_2}{\varphi_2}
					}
				\end{equation}
			\end{minipage}
		&	\begin{minipage}{\WIDTH}
				\noindent%
				\begin{equation}
					\label{rule:triplprovestot+seq}
					\dfracx[r]{%
						\triplprovestot \tripl{\varphi_1}{\pi_1}{\varphi}
					\\	\LAND \triplprovestot \tripl{\varphi}{\pi_2}{\varphi_2}
					}{%
						\triplprovestot \tripl{\varphi_1}{\pi_1 \commseq \pi_2}{\varphi_2}
					}
				\end{equation}
			\end{minipage}
		\\
		\\	\begin{minipage}{\WIDTH}
				\noindent%
				\begin{equation}
					\label{rule:triplproves+cons}
					\dfracx[l]{%
						\triplproves \tripl{\varphi_1'}{\pi}{\varphi_2'}
					\\	\LAND \varphi_1 \Rightarrow \varphi_1' \AND \varphi_2' \Rightarrow \varphi_2
					}{%
						\triplproves \tripl{\varphi_1}{\pi}{\varphi_2}
					}
				\end{equation}
			\end{minipage}
		&	\begin{minipage}{\WIDTH}
				\noindent%
				\begin{equation}
					\label{rule:triplprovestot+cons}
					\dfracx[l]{%
						\triplprovestot \tripl{\varphi_1'}{\pi}{\varphi_2'}
					\\	\LAND \varphi_1 \Rightarrow \varphi_1' \AND \varphi_2' \Rightarrow \varphi_2
					}{%
						\triplprovestot \tripl{\varphi_1}{\pi}{\varphi_2}
					}
				\end{equation}
			\end{minipage}
		\\
		\\	\begin{minipage}{\WIDTH}
				\noindent%
				\begin{equation}
					\label{rule:triplproves+fail}
					\dfracx{%
						\triplproves \tripl{\varphi \wedge \ell}{\pi}{\varphi'}
					}{%
						\triplproves \tripl{\varphi}{\commifthen{\ell}{P}}{\varphi'}
					}
				\end{equation}
			\end{minipage}
		&	\begin{minipage}{\WIDTH}
				\noindent%
				\begin{equation}
					\label{rule:triplprovestot+fail}
					\dfracx[l]{%
						\triplprovestot \tripl{\varphi}{\pi}{\varphi'} \AND \varphi \Rightarrow \ell
					}{%
						\triplprovestot \tripl{\varphi}{\commifthen{\ell}{\pi}}{\varphi'}
					}
				\end{equation}
			\end{minipage}
		\\	
		\\	\multicolumn{2}{@{} c @{}}{%
				\begin{minipage}{\linewidth}
					\noindent%
					\begin{equation}
						\label{rule:triplprovestot+decomp}
						\dfracx[r]{%
							\triplproves \tripl{\varphi}{\pi}{\varphi_1} \AND \triplprovestot \tripl{\varphi}{\pi}{\varphi_2}
						}{%
							\triplprovestot \tripl{\varphi}{\pi}{\varphi_1 \wedge \varphi_2}
						}
					\end{equation}
				\end{minipage}
			}
		\end{tabular}
	\end{framed}
	
	\caption{Addendum to Definition~\ref{def:triplproves}}
	\label{fig:def+triplproves}
\end{figure}

\begin{defi}
	[proof systems of triples]
	\label{def:triplproves}
	\label{def:triplprovestot}
	${\triplproves} \, {\triplprovestot} \subseteq \tripluniv$ denote the smallest relations induced by the rules in Figure~\ref{fig:def+triplproves}.
\end{defi}
\newpage
\begin{prop}
	[Theorem 3.6 in {\cite[Section 3.7]{AdBO09}}]
	\label{thm:triplproves+sound}
	\strut
	\begin{itemize}
		\item
		$\triplproves \tripl{\varphi}{\pi}{\varphi'} \IMPLIES \triplmodels \tripl{\varphi}{\pi}{\varphi'}$
		
		\item
		$\triplprovestot \tripl{\varphi}{\pi}{\varphi'} \IMPLIES \triplmodelstot \tripl{\varphi}{\pi}{\varphi'}$
	\end{itemize}
\end{prop}

\noindent
Note that the first four rules for~$\triplproves$ and the first four rules for~$\triplprovestot$ have the same premise/consequence.
We use~$\triplproves$ to prove the \emph{soundness} of commandification; We use~$\triplprovestot$ to prove commandification's \emph{completeness}.

\subsection*{Commandification (without Cycles)}

At run-time, to check if a transition~$\tpl{q \, P \, \varphi \, q'}$ can fire, a \compilergenerated coordinator thread first checks every port in~$P$ \edit{for \emph{readiness}}.
For instance, every (data structure for an) input port should have a pending \code{put}.
Subsequently, the coordinator thread checks whether a data state~$\sigma$ exists that (i) satisfies~$\varphi$ and (ii) subsumes an \emph{initial data state}~$\sigmai$ (i.e.,~$\sigmai \subseteq \sigma$).
If so, we call~$\sigma$ a \emph{solution} of~$\varphi$ under~$\sigmai$.
The domain of~$\sigmai$ contains all \emph{uncontrollable data variables} in~$\varphi$: the input ports in~$P$ (intersected with~$\freefun(\varphi)$) and~$\pre{m}$ for every memory cell~$m$ in the \CA (also intersected with~$\freefun(\varphi)$).
More precisely,~$\sigmai$ maps every input port~$p$ in~$\freefun(\varphi)$ to the particular datum \emph{forced} to pass through~$p$ by the process thread on \editt{the other side of~$p$} (i.e., the datum involved in~$p$'s pending \code{put}), while~$\sigmai$ maps every~$\pre{m}$ in~$\freefun(\varphi)$ to the datum that currently resides in~$m$.
Thus, before the coordinator thread invokes a constraint solver for~$\varphi$, it already fixes values for all uncontrollable data variables in~$\varphi$; when subsequently invoked, a constraint solver may, in search of a solution for~$\varphi$ under~$\sigmai$, select values only for data variables outside~$\sigmai$'s domain.
Slightly more formally:
$$
\label{page:deltai}
\begin{array}{@{} l @{\ } c @{\ } l @{}}
	\sigmai	& =	& \setbx[l][l]{p \mapsto d}{%
					\SCOPE{\text{the \code{put} pending on input port } p \text{ involves datum } d}
				\\	\LAND p \in \freefun(\varphi)
				}
\\	\UP
\\			&	& {} \cup \setbx[l][l]{\pre{m} \mapsto d}{%
					\SCOPE{\text{memory cell } m \text{ currently contains datum } d}
				\\	\LAND \pre{m} \in \freefun(\varphi)
				}
\end{array}
$$
With commandification, instead of invoking a constraint solver, the coordinator thread executes a \compilergenerated data command for~$\varphi$ on~$\sigmai$, thereby gradually extending~$\sigmai$ to a full solution.
This \compilergenerated data command essentially works as an efficient, \editt{small}, dedicated constraint solver for~$\varphi$.

To translate a data constraint of the form~$\ell_1 \wedge \cdots \wedge \ell_k$, we construct a data command that (i) enforces as many data literals of the form~$t_1 \termeq t_2$ as possible with \editt{assignment statements} and (ii) checks all remaining data literals with failure statements.
We call data literals of the form~$t_1 \termeq t_2$ \emph{data equalities}.
To examplify such commandification, recall data constraint~$\phieg$ on page~\pageref{page:phieg}.
In this data constraint, let~$\reo{C}$ denote an input port and let~$\reo{x}$ denote a memory cell.
In that case, the set of uncontrollable data variables in~$\phieg$ consists of~$\reo{C}$ and~$\pre{\reo{x}}$.
Now,~$\phieg$ has six correct commandifications:
$$
\LINES{%
	\pi_1 = \LINES[l][t]{%
		\reo{B} \commasgn {\pre{\reo{x}}} \commseq {}
	\\	\reo{D} \commasgn \reo{C} \commseq {}
	\\	\reo{E} \commasgn \code{add}(\reo{B} \, \reo{D}) \commseq {}
	\\	\reo{F} \commasgn \reo{E} \commseq {}
	\\	\reo{G} \commasgn \reo{E}
	\\	\commifthen{\neg \code{Odd}(\reo{G})}{\commskip} \commseq {} 
	} \quad \pi_2 = \LINES[l][t]{%
		\reo{B} \commasgn \pre{\reo{x}} \commseq {}
	\\	\reo{D} \commasgn \reo{C} \commseq {}
	\\	\reo{E} \commasgn \code{add}(\reo{B} \, \reo{D}) \commseq {}
	\\	\reo{G} \commasgn \reo{E}
	\\	\reo{F} \commasgn \reo{E} \commseq {}
	\\	\commifthen{\neg \code{Odd}(\reo{G})}{\commskip} \commseq {} 
	} \quad \pi_3 = \LINES[l][t]{%
		\reo{B} \commasgn \pre{\reo{x}} \commseq {}
	\\	\reo{D} \commasgn \reo{C} \commseq {}
	\\	\reo{E} \commasgn \code{add}(\reo{B} \, \reo{D}) \commseq {}
	\\	\reo{G} \commasgn \reo{E}
	\\	\commifthen{\neg \code{Odd}(\reo{G})}{\commskip} \commseq {} 
	\\	\reo{F} \commasgn \reo{E} \commseq {}
	}
\\	
\\	\pi_4 = \LINES[l][t]{%
		\reo{D} \commasgn \reo{C} \commseq {}
	\\	\reo{B} \commasgn \pre{\reo{x}} \commseq {}
	\\	\reo{E} \commasgn \code{add}(\reo{B} \, \reo{D}) \commseq {}
	\\	\reo{F} \commasgn \reo{E} \commseq {}
	\\	\reo{G} \commasgn \reo{E}
	\\	\commifthen{\neg \code{Odd}(\reo{G})}{\commskip} \commseq {} 
	} \quad \pi_5 = \LINES[l][t]{%
		\reo{D} \commasgn \reo{C} \commseq {}
	\\	\reo{B} \commasgn \pre{\reo{x}} \commseq {}
	\\	\reo{E} \commasgn \code{add}(\reo{B} \, \reo{D}) \commseq {}
	\\	\reo{G} \commasgn \reo{E}
	\\	\reo{F} \commasgn \reo{E} \commseq {}
	\\	\commifthen{\neg \code{Odd}(\reo{G})}{\commskip} \commseq {} 
	} \quad \pi_6 = \LINES[l][t]{%
		\reo{D} \commasgn \reo{C} \commseq {}
	\\	\reo{B} \commasgn \pre{\reo{x}} \commseq {}
	\\	\reo{E} \commasgn \code{add}(\reo{B} \, \reo{D}) \commseq {}
	\\	\reo{G} \commasgn \reo{E}
	\\	\commifthen{\neg \code{Odd}(\reo{G})}{\commskip} \commseq {} 
	\\	\reo{F} \commasgn \reo{E} \commseq {}
	}
}
$$
We stipulate the same precondition for each of these data commands, namely that~$\pre{\reo{x}}$ and~$\reo{C}$ have a non-$\nil$ value (later formalized as data literals~$\pre{\reo{x}} \termeq \pre{\reo{x}}$ and~$\reo{C} \termeq \reo{C}$).
This precondition models that the execution of these data commands should always start on an initial data state over the uncontrollable data variables~$\pre{\reo{x}}$ and~$\reo{C}$.
Under this precondition, if a coordinator thread executes~$\pi_1$, it first assigns the values of~$\pre{\reo{x}}$ and~$\reo{C}$ to~$\reo{B}$ and~$\reo{D}$.
Subsequently, it assigns the evaluation of~$\code{add}(\reo{B} \, \reo{D})$ to~$\reo{E}$.
Next, it assigns the value of~$\reo{E}$ to~$\reo{F}$ and~$\reo{G}$.
Finally, it checks~$\neg \code{Odd}(\reo{G})$ with a failure statement.
Data commands~$\pi_2$ and~$\pi_3$ differ from data command~$\pi_1$ only in the order of the last three steps; data commands~$\pi_4$,~$\pi_5$ and~$\pi_6$ differ from~$\pi_1$,~$\pi_2$ and~$\pi_3$ only in the order of the first two steps.
If execution of~$\pi_i$ on~$\sigmai$ successfully terminates, the resulting final data state~$\sigma$ satisfies~$\phieg$.
We call this \emph{soundness}.
Moreover, if a~$\sigma'$ exists such that~$\sigma' \dcmodels \phieg$ and~$\sigmai \subseteq \sigma'$, execution of~$\pi_i$ successfully terminates.
We call this \emph{completeness}.

Generally, soundness and completeness crucially depend on the order in which assignments and failure statements follow each other in~$\pi$.
For instance, changing the order of~$\reo{G} \commasgn \reo{E}$ and~$\commifthen{\neg \code{Odd}(\reo{G})}{\commskip}$ in the previous example yields a data command whose execution always fails (because~$\reo{G}$ does not have a value yet on evaluating the guard of the failure statement).
Such a trivially sound but incomplete data constraint serves no \editt{purpose}.
As another complication, not every data equality can become an assignment.
In a first class of cases, neither the left-hand side nor the right-hand side of a data equality matches data variable~$x$.
For instance, We \emph{must} translate~$\code{add}(\reo{B} \, \reo{D}) \termeq \code{mult}(\reo{B} \, \reo{D})$ into a failure statement, because we clearly cannot assign either of its two operands to the other.
In a second class of cases, multiple data equalities in a data constraint have a left-hand side or a right-hand side that matches the same data variable~$x$.
For instance, we can translate only one data equality in~$\reo{E} \termeq \code{add}(\reo{B} \, \reo{D}) \wedge \reo{E} \termeq \code{mult}(\reo{B} \, \reo{D})$ into an assignment, after which we \emph{must} translate the other one into a failure statement, to avoid conflicting assignments to~$\reo{E}$.

To deal with these complications, we define a \emph{precedence relation} on the data literals in a data constraint that formalizes their dependencies.
Recall from Definition~\ref{def:dcuniv} that every data constraint consists of a conjunctive kernel of data literals, enveloped with existential quantifications.
First, for technical convenience, we introduce a function that extends~$\litfun(\varphi)$ (i.e., the data literals in the kernel of~$\varphi$) with ``symmetric data equalities''.

\begin{defi}
	[$\termeq$-symmetric closure]
	\label{def:symlitfun}
	$\symlitfun : \dcuniv \rightarrow \pset{\dcuniv}$ denotes the function defined by the following equation:
	$$
	\symlitfun(\varphi) = \litfun(\varphi) \cup \setb{t_2 \termeq t_1}{t_1 \termeq t_2 \in \litfun(\varphi)}
	$$
\end{defi}

\noindent
Obviously, because~$t_1 \termeq t_2 \dcequiv t_2 \termeq t_1$, we have~$\bigwedge \litfun(\varphi) \dcequiv \bigwedge \symlitfun(\varphi)$ for all~$\varphi$.

\begin{figure}
	\centering
	
	\renewcommand{\WIDTH}{.5\linewidth}
	\begin{framed}
		\begin{tabular}{@{}c@{}c@{}}
			\begin{minipage}{\WIDTH}
				\noindent%
				\begin{equation}
					\label{rule:prec+free}
					\hphantom{x} \dfracx[l]{%
						x \termeq t \, \ell \in \symlitfun(\varphi)
					\\	\LAND x \in \variablfun(\ell)
					}{%
						x \termeq t \prec \ell
					} \hphantom{x}
				\end{equation}
			\end{minipage}
		&	\begin{minipage}{\WIDTH}
				\noindent%
				\begin{equation}
					\label{rule:prec+conv}
					\dfracx[l]{%
						x \termeq t \, \ell \in \symlitfun(\varphi)
					\\	\LAND \SCOPE{\ell \neq x' \termeq t' \FORALL x' \, t'}
					}{%
						x \termeq t \prec \ell
					}
				\end{equation}
			\end{minipage}
		\\
		\\	\multicolumn{2}{@{}c@{}}{%
				\begin{minipage}{\linewidth}
					\noindent%
					\begin{equation}
						\label{rule:prec+trans}
						 \dfrac{%
							\ell_1 \prec \ell_2 \AND \ell_2 \prec \ell_3 \AND \ell_2 \notin \set{\ell_1 \, \ell_3}
						}{%
							\ell_1 \prec \ell_3
						}
					\end{equation}
				\end{minipage}
			}
		\end{tabular}
	\end{framed}
	
	\caption{Addendum to Definition~\ref{def:prec}}
	\label{fig:def+prec1}
\end{figure}

\begin{defi}
	[precedence \I]
	\label{def:prec}
	${\prec} : \dcuniv \rightarrow \pset{\dcuniv \times \dcuniv}$ denotes the function defined by the following equation:
	$$
	{\prec}(\varphi) = {\prec}
	$$
	where~$\prec$ denotes the smallest relation induced by the rules in Figure~\ref{fig:def+prec1}.
\end{defi}

We usually write~$\prec_\varphi$ instead of~${\prec}(\varphi)$ and use~$\prec_\varphi$ as an infix relation.
In words,~$x \termeq t \prec_\varphi \ell$ means that the assignment~$x \commasgn t$ precedes the commandification of~$\ell$ (i.e.,~$\ell$ depends on~$x$).
Rule~\ref{rule:prec+free} deals with the previously discussed first class of da\-ta-e\-qual\-i\-ties-that-can\-not-be\-come-as\-sign\-ments, by imposing precedence only on data literals of the form~$x \termeq t$; shortly, we comment on the second class of da\-ta-e\-qual\-i\-ties-that-can\-not-be\-come-as\-sign\-ments.
Rule~\ref{rule:prec+conv} conveniently ensures that every~$x \termeq t$ precedes all differently shaped data literals. 
Strictly speaking, we do not need this rule, but it simplifies some notation and proofs later on.

\begin{figure}[t]
	\centering
	\renewcommand{\reosize}{\tiny}
	\begin{tikzpicture}
		\newcommand{\vertex}[4]{%
			\node[draw, fill=lightgray, circle, inner sep=1.5pt, outer sep=1pt] (#1) at (#2:4cm) {};
			\draw[out=#2-1, in=#2+1, min distance=0pt, draw=white] (#1) to node [sloped, #3] {\scriptsize#4} (#1);
		}
		
		\newcommand{\arc}[3][]{%
			\draw[->, bend left=0, lightgray, #1]	(#2) to (#3);
		}
		
		\newcommand{\fatarc}[3][]{%
			\draw[->, bend left=0, white, line width=3pt, #1]	(#2) to (#3);
			\draw[->, bend left=0, black, line width=2pt, #1]	(#2) to (#3);
		}
		
		\vertex{B}		{ 5*180/7}{above}{$\pre{\reo{x}} \termeq \reo{B}$}
		\vertex{D}		{ 4*180/7}{above}{$\reo{C} \termeq \reo{D}$}
		\vertex{E}		{ 3*180/7}{above}{$\code{add}(\reo{B} \, \reo{D}) \termeq \reo{E}$}
		\vertex{F}		{ 2*180/7}{above}{$\reo{E} \termeq \reo{F}$}
		\vertex{G}		{ 1*180/7}{above}{$\reo{E} \termeq \reo{G}$}
		
		\vertex{B'}		{-5*180/7}{below}{$\reo{B} \termeq \pre{\reo{x}}$}
		\vertex{D'}		{-4*180/7}{below}{$\reo{D} \termeq \reo{C}$}
		\vertex{E'}		{-3*180/7}{below}{$\reo{E} \termeq \code{add}(\reo{B} \, \reo{D})$}
		\vertex{F'}		{-2*180/7}{below}{$\reo{F} \termeq \reo{E}$}
		\vertex{G'}		{-1*180/7}{below}{$\reo{G} \termeq \reo{E}$}
		
		\vertex{Odd}	{ 0*180/7}{below}{$\neg \code{Odd}(\reo{G})$}
		
		\arc{B}		{B'}
		\arc{D}		{D'}
		\arc{F}		{E}
		\arc{F}		{G}
		\arc{F}		{E'}
		\arc{F}		{F'}
		\arc{F}		{G'}
		\arc{G}		{E}
		\arc{G}		{F}
		\arc{G}		{E'}
		\arc{G}		{F'}
		\arc{G}		{G'}
		
		\arc{B'}	{B}
		\arc{B'}	{E}
		\arc{D'}	{D}
		\arc{D'}	{E}
		\arc{E'}	{E}
		\arc{E'}	{F}
		\arc{E'}	{G}
		\arc{F'}	{F}
		\arc{G'}	{G}
		
		\fatarc{B'}	{E'}
		\fatarc{D'}	{E'}
		\fatarc{E'}	{F'}
		\fatarc{E'}	{G'}
		\fatarc{G'}	{Odd}
	\end{tikzpicture}
	
	\caption{%
		Digraph for precedence relation~$\prec_\phieg$ (without loop arcs and without arcs induced by Rule~\ref{rule:prec+conv}, to avoid further clutter).
		An arc~$\tpl{\ell \, \ell'}$ corresponds to~$\ell \prec_\phieg \ell'$.
		Arcs between the same data vertices, but in different directions, lie on top of each other.
		Bold arcs represent a fragment of the strict partial order extracted from~$\prec_\phieg$.
	}
	\label{fig:prec}
\end{figure}

For the sake of argument---generally, this does \emph{not} hold true---suppose that a precedence relation~$\prec_\varphi$ denotes a \emph{strict partial order} on~$\symlitfun(\varphi)$.
In that case, we can \emph{linearize}~$\prec_\varphi$ to a strict total order~$\prectotal$ (i.e., embedding~$\prec_\varphi$ into~$\prectotal$ such that~${\prec_\varphi} \subseteq {\prectotal}$) with a topological sort on the digraph~$\tpl{\symlitfun(\varphi) \, \prec_\varphi}$~\cite{Kah62,Knu97}.
Intuitively, such a linearization gives us an order in which we can translate data literals in~$\symlitfun(\varphi)$ to data commands in a sound and complete way.
Shortly, we give an algorithm for doing so and indeed prove its correctness.
Problematically, however,~$\prec_\varphi$ generally does not denote a strict partial order: generally, it violates asymmetry and irreflexivity (i.e., graph-theoretically, it contains many cycles).
For instance, Figure~\ref{fig:prec} shows the digraph~$\tpl{\symlitfun(\phieg) \, \prec_\phieg}$, which indeed contains cycles.
For now, we defer this issue to the next subsection, because it forms a concern orthogonal to the commandification algorithm and its correctness.
Until then, we simply assume the existence of a procedure for extracting a strict partial order from~$\prec_\varphi$, represented by bold arcs in Figure~\ref{fig:prec}.

\begin{algorithm}[t]
	\begin{algorithmic}
		\Require {$\LINES[l][t]{%
			{\prectotal} \text{ denotes a strict total order on } \symlitfun(\varphi)
		\\	\LAND \symlitfun(\varphi) = \set{\ell_1 \, \ldots \, \ell_{n+m}}
		\\	\LAND \ell_1 \prectotal \cdots \prectotal \ell_n \prectotal \ell_{n+1} \prectotal \cdots \prectotal \ell_{n+m}
		\\	\LAND \ell_1 = x_1 \termeq t_1 \AND \cdots \AND \ell_n = x_n \termeq t_n
		\\	\LAND \variablfun(\varphi) \setminus X \subseteq \set{x_1 \, \ldots \, x_n}
		\\	\UP
		\\	\LAND \SCOPE{%
				\SCOPEX{%
					\SCOPE{x \termeq t \in \symlitfun(\varphi) \AND x' \in \variablfun(t)} \RIMPLIES
				\\	\SCOPE{x' \in X \OR \SCOPE{x' \termeq t' \prectotal x \termeq t \FORSOME t'}}
				} \FORALL x \, x' \, t
			}
		}$}
		
		\State \vphantom{foo}
		\Function {\algor{commandification}}{$\varphi \, X \, \prectotal$}
			\State {$\pi := \commskip$}
			\State {$i := 1$}
			\While {$i \leq n$}
				\If {$x_i \in X \cup \set{x_1 \, \ldots \, x_{i-1}}$}
					\State {$\pi := (\pi \commseq \commifthen{x_i \termeq t_i}{\commskip})$}
				\Else
					\State {$\pi := (\pi \commseq x_i \commasgn t_i)$}
				\EndIf
				\State {$i := i + 1$}
			\EndWhile
			\While {$i \leq n + m$}
				\State {$\pi := (\pi \commseq \commifthen{\ell_i}{\commskip})$}
				\State {$i := i + 1$}
			\EndWhile 
			\State \Return {$\pi$}
		\EndFunction
		\State \vphantom{foo}
		
		\Ensure {$\LINES[l][t]{%
			\triplproves \tripl{\bigwedge \setb{x \termeq x}{x \in X}}{\pi}{\ell_1 \wedge \cdots \wedge \ell_{n+m}}
		\\	\UP
		\\	\LAND \SCOPE{\SCOPEX{%
				\sigma \dcmodels \ell_1 \wedge \cdots \wedge \ell_{n+m} \RIMPLIES 
			\\	\UP
			\\	\triplprovestot \triplx{\bigwedge \setb{x \termeq \sigma(x)}{x \in X}}{\pi}{\bigwedge \setb{x \termeq \sigma(x)}{x \in X \cup \set{x_1 \, \ldots \, x_n}}}
			} \FORALL \sigma}
		}$}
	\end{algorithmic}
	
	\caption{Algorithm for translating a data constraint~$\varphi$, a set of data variables~$X$, and a binary relation on data literals~$\prectotal$ to a data command~$\pi$}
	\label{algo:commandification}
\end{algorithm}

Algorithm~\ref{algo:commandification} translates a data constraint~$\varphi$, a set of data variables~$X$, and a binary relation on data literals~$\prectotal$ to a data command~$\pi$.
It \textbf{require}s the following on its input.
First,~$\prectotal$ should denote a strict total order on the~$\termeq$-symmetric closure of~$\varphi$'s data literals.
Let~$n$ denote \emph{a}---not necessarily \emph{the}---number of data equalities in~$\symlitfun(\varphi)$, and let~$m$ denote the number of remaining data literals in~$\symlitfun(\varphi)$.
Then,~$\ell_1 \, \ldots \, \ell_{n+m}$ denote the data literals in~$\symlitfun(\varphi)$ such that (i) their indices respect~$\prectotal$ and (ii) every~$\ell_i$ denotes~$x_i \termeq t_i$ for~$1 \leq i \leq n$.
Next, for every data variable in a data literal in~$\symlitfun(\varphi)$, but outside the set of uncontrollable data variables~$X$, a data equality~$x_i \termeq t_i$ should exist.
Otherwise, such a data variable can get a value only through search---exactly what commandification tries to avoid---and not through assignment; \emph{underspecified} data constraints fundamentally lie outside the scope of commandification in general and Algorithm~\ref{algo:commandification} in particular.
Finally, if \editt{a} term~$t$ in a data equality~$x \termeq t$ depends on a variable~$x'$, a data equality~$x' \termeq t'$ should precede~$x \termeq t$ under~$\prectotal$.
The rules in Definition~\ref{def:prec} induce precedence relations for which all these \textbf{require}ments hold true, except that those precedence relations \editt{do} not necessarily denote strict partial orders and, hence, may not admit linearization.
Consequently, the precedence relations in Definition~\ref{def:prec} may not yield strict total orders as \textbf{require}d by Algorithm~\ref{algo:commandification}.
We address this issue in the next subsection.

Assuming satisfaction of its \textbf{require}ments, Algorithm~\ref{algo:commandification} works as follows.
It first loops over the first~$n$ (according to~$\prectotal$)~$x_i \termeq t_i$ data literals.
If an assignment for~$x_i$ already exists in the data command under construction~$\pi$, Algorithm~\ref{algo:commandification} translates~$x_i \termeq t_i$ to a failure statement; otherwise, it translates~$x_i \termeq t_i$ to an assignment.
This approach resolves issues with the previously discussed second class of e\-qual\-i\-ties-that-can\-not-be\-come-as\-sign\-ments.
After the first loop, the algorithm uses a second loop to translate the remaining~$m$ data literals to failure statements.
The algorithm runs in time linear in~$n + m$, and it terminates.

Upon termination, Algorithm~\ref{algo:commandification} \textbf{ensure}s the soundness (first conjunct) and completeness of~$\pi$ (second conjunct).
Note that we use a different proof system for soundness (partial correctness,~$\triplproves$) than for completeness (total correctness,~$\triplprovestot$).

\begin{thm}
	\label{thm:algo+commandification}
	Algorithm~\ref{algo:commandification} is correct.
\end{thm}

Algorithm~\ref{algo:commandification} has the minor issue that it may produce more failure statements than strictly necessary.
For instance, if we run Algorithm~\ref{algo:commandification} on the total order extracted from~$\prec_\phieg$ in Figure~\ref{fig:prec}, we get both the assignment~$\reo{D} \commasgn \reo{C}$ and the unnecessary failure statement~$\commifthen{\reo{C} \termeq \reo{D}}{\commskip}$.
After all, the digraph contains both~$\reo{D} \termeq \reo{C}$ and~$\reo{C} \termeq \reo{D}$, one of which we added while computing~$\symlitfun(\phieg)$ to account for the symmetry of~$\termeq$.
Generally, such symmetric data literals result either in one assignment and one failure statement or in two failure statements; one can easily prove that symmetric data literals never result in two assignments.
In both cases, one can safely remove one of the failure statements, because successful termination of the remaining statement already accounts for the removed failure statement.

\subsection*{Commandification (with Cycles)}

Algorithm~\ref{algo:commandification} \textbf{require}s that~$\prectotal$ denotes a strict total order.
Precedence relations in Definition~\ref{def:prec} of~$\prec$, however, do not yield such orders: graph-theoretically, they may contain cycles.
In this subsection, we present a solution for this problem.
We start by extending the previous precedence relations with a unique least element,~$\bigstar$, and by making dependencies of data literals on uncontrollable data variables explicit.
In the following definition, let~$X$ denote a set of such variables.

\begin{figure}
	\centering
	
	\renewcommand{\WIDTH}{.33\linewidth}
	\begin{framed}
		\begin{tabular}{@{}c@{}c@{}c@{}}
			\begin{minipage}{\WIDTH}
				\noindent%
				\begin{equation}
					\dfracx{%
						\strut
					\\	\ell_1 \prec_\varphi \ell_2
					}{%
						\ell_1 \precx \ell_2
					}
				\end{equation}
			\end{minipage}
		&	\begin{minipage}{\WIDTH}
				\noindent%
				\begin{equation}
					\dfracx[l]{%
						\ell \in \symlitfun(\varphi)
					\\	\LAND \variablfun(\ell) \subseteq X
					}{%
						\bigstar \precx \ell
					}
				\end{equation}
			\end{minipage}
		&	\begin{minipage}{\WIDTH}
				\noindent%
				\begin{equation}
					\dfracx[l]{%
						x \termeq t \in \symlitfun
					\\	\LAND \variablfun(t) \subseteq X
					}{%
						\bigstar \precx x \termeq t
					}
				\end{equation}
			\end{minipage}
		\end{tabular}
	\end{framed}
	
	\caption{Addendum to Definition~\ref{def:precx}}
	\label{fig:def+precx}
\end{figure}

\begin{defi}
	[precedence \II]
	\label{def:precx}
	${\precx} : \dcuniv \times \pset{\xuniv} \rightarrow \pset{(\dcuniv \cup \set{\bigstar}) \times \dcuniv}$ denotes the function defined by the following equation:
	$$
	{\precx}(\varphi \, X) = {\precx}
	$$
	where~$\precx$ denotes the smallest relation induced by the rules in Figure~\ref{fig:def+precx}.
\end{defi}

We usually write~$\prec_\varphi^X$ instead of~${\prec}(\varphi \, X)$ and use~$\prec_\varphi^X$ as an infix relation.
The two new rules state that data literals in which only uncontrollable data variables occur ``depend'' on~$\bigstar$.

Relation~$\prec_\varphi^X$ denotes a strict partial order if its digraph~$\tpl{\symlitfun(\varphi) \cup \set{\bigstar} \, \prec_\varphi^X}$ defines a~$\bigstar$-\emph{arborescence}: a digraph consisting of~$n - 1$ arcs such that a path exists from~$\bigstar$ to each of its~$n$ vertices~\cite{KV08}.
Equivalently, in a~$\bigstar$-arborescence,~$\bigstar$ has no incoming arcs, every other vertex has exactly one incoming arc, and the arcs form no cycles~\cite{KV08}.
The first formulation seems more intuitive here: every path from~$\bigstar$ to some data literal~$\ell$ represents an order in which Algorithm~\ref{algo:commandification} should translate the data literals on that path to ensure the correctness of the translation of~$\ell$.
The second formulation simplifies observing that arborescences correspond to strict partial orders.

\begin{figure}[t]
	\centering
	\renewcommand{\reosize}{\tiny}
	\begin{tikzpicture}
		\newcommand{\vertex}[4]{%
			\node[draw, fill=lightgray, circle, inner sep=1.5pt, outer sep=1pt] (#1) at (#2:4cm) {};
			\draw[out=#2-1, in=#2+1, min distance=0pt, draw=white] (#1) to node [sloped, #3] {\scriptsize#4} (#1);
		}
		
		\newcommand{\arc}[3][]{%
			\draw[->, bend left=0, lightgray, #1]	(#2) to (#3);
		}
		
		\newcommand{\fatarc}[3][]{%
			\draw[->, bend left=0, white, line width=3pt, #1]	(#2) to (#3);
			\draw[->, bend left=0, black, line width=2pt, #1]	(#2) to (#3);
		}
		
		\newcommand{\biarc}		[4][]{%
			\tkzCentroid(#2,#3,#4)
			\tkzGetPoint{Centroid}
			\tkzDefMidPoint(Centroid,#4)
			\tkzGetPoint{MidPoint}
			\tkzDefMidPoint(Centroid,MidPoint)
			\tkzGetPoint{MidPointMidPoint}
			\draw [bend left=0, lightgray, #1] (#2) to (MidPointMidPoint);
			\draw [bend left=0, lightgray, #1] (#3) to (MidPointMidPoint);
			\draw [->, lightgray, #1] (MidPointMidPoint) to (#4);
		}
		
		\newcommand{\fatbiarc}		[4][]{%
			\tkzCentroid(#2,#3,#4)
			\tkzGetPoint{Centroid}
			\tkzDefMidPoint(Centroid,#4)
			\tkzGetPoint{MidPoint}
			\tkzDefMidPoint(Centroid,MidPoint)
			\tkzGetPoint{MidPointMidPoint}
			\draw [bend left=0, white, line width=3pt, #1] (#2) to (MidPointMidPoint);
			\draw [bend left=0, white, line width=3pt, #1] (#3) to (MidPointMidPoint);
			\draw [->, white, line width=3pt, #1] (MidPointMidPoint) to (#4);
			
			\draw [bend left=0, black, line width=2pt, #1] (#2) to (MidPointMidPoint);
			\draw [bend left=0, black, line width=2pt, #1] (#3) to (MidPointMidPoint);
			\draw [->, black, line width=2pt, #1] (MidPointMidPoint) to (#4);
		}
		
		\vertex{Init}	{ 7*180/7}{below}{$\bigstar$}
		
		\vertex{x}		{ 6*180/7}{above}{$\pre{\reo{x}} \termeq \pre{\reo{x}}$}
		\vertex{B}		{ 5*180/7}{above}{$\pre{\reo{x}} \termeq \reo{B}$}
		\vertex{D}		{ 4*180/7}{above}{$\reo{C} \termeq \reo{D}$}
		\vertex{E}		{ 3*180/7}{above}{$\code{add}(\reo{B} \, \reo{D}) \termeq \reo{E}$}
		\vertex{F}		{ 2*180/7}{above}{$\reo{E} \termeq \reo{F}$}
		\vertex{G}		{ 1*180/7}{above}{$\reo{E} \termeq \reo{G}$}
		
		\vertex{C}		{-6*180/7}{below}{$\reo{C} \termeq \reo{C}$}
		\vertex{B'}		{-5*180/7}{below}{$\reo{B} \termeq \pre{\reo{x}}$}
		\vertex{D'}		{-4*180/7}{below}{$\reo{D} \termeq \reo{C}$}
		\vertex{E'}		{-3*180/7}{below}{$\reo{E} \termeq \code{add}(\reo{B} \, \reo{D})$}
		\vertex{F'}		{-2*180/7}{below}{$\reo{F} \termeq \reo{E}$}
		\vertex{G'}		{-1*180/7}{below}{$\reo{G} \termeq \reo{E}$}
		
		\vertex{Odd}	{ 0*180/7}{below}{$\neg \code{Odd}(\reo{G})$}
		
		\arc	{B}			{x}
		\arc	{D}			{C}
		\biarc	{C}{D'}		{D}
		\biarc	{G}{F'}		{F}
		\arc	{F'}		{F}
		\biarc	{F}{G'}		{G}
		\arc	{G'}		{G}
		
		\arc	{x}			{B'}
		\arc	{B}			{B'}
		\arc	{D}			{D'}
		\arc	{F}			{F'}
		\arc	{G}			{F'}
		\arc	{F}			{G'}
		\arc	{G}			{G'}
		
		\fatarc		{Init}		{x}
		\fatarc		{x}			{B'}
		\fatbiarc	{x}{B'}		{B}
		\fatarc		{Init}		{C}
		\fatarc		{C}			{D'}
		\fatarc		{D'}		{D}
		\fatbiarc	{B'}{D'}	{E'}
		\fatarc		{E'}		{F'}
		\fatbiarc	{E'}{F'}	{F}
		\fatarc		{E'}		{G'}
		\fatbiarc	{E'}{G'}	{G}
		\fatarc		{G'}		{Odd}
	\end{tikzpicture}
	
	\caption{%
		B-graph corresponding to the digraph in Figure~\ref{fig:prec} (without loop \barcs and without three-tailed \barcs, to avoid further clutter).
		An arc~$\tpl{\ell \, \ell'}$ corresponds to~$\ell \prec_\phieg \ell'$.
		Bold arcs represent an arborescence.
	}
	\label{fig:bprec}
\end{figure}

A naive approach to extract a strict partial order from~$\prec_\varphi^X$ consists of computing a~$\bigstar$-arborescence of the digraph~$\tpl{\symlitfun(\varphi) \cup \set{\bigstar} \, \prec_\varphi^X}$.
Even if such a~$\bigstar$-arborescence exists, however, this approach does not work as expected if~$\symlitfun(\varphi)$ contains a data literal~$x \termeq t$ where~$t$ has more than one data variable.
For instance, by definition, every arborescence of the digraph in Figure~\ref{fig:prec} has only one incoming arc for~$\reo{E} \termeq \code{add}(\reo{B} \, \reo{D})$, even though assignments to \emph{both}~$\reo{B}$ \emph{and}~$\reo{D}$ must precede an assignment to~$\reo{E}$.
Because these dependencies exist as two separate arcs, no arborescence can capture them.
To solve this, we \editt{must} somehow represent the dependencies of~$\reo{E} \termeq \code{add}(\reo{B} \, \reo{D})$ with a single incoming arc.
We can do so by allowing arcs to have multiple tails, one for every data variable.
In that case, we can replace the two separate incoming arcs of~$\reo{E} \termeq \code{add}(\reo{B} \, \reo{D})$ with a single two-tailed incoming arc as in Figure~\ref{fig:bprec}.
The two tails make explicit that to evaluate~$\code{add}$, we need values for both its arguments: multiple tails represent a conjunction of dependencies of a data literal.

By combining single-tailed arcs into multiple-tailed arcs, we effectively transform the digraphs considered so far into \emph{\bgraphs}, a special kind of hypergraph with only \emph{\barcs} (i.e., \emph{b}ackward hyperarcs, i.e., hyperarcs with exactly one head)~\cite{GLPN93}.
Generally, we cannot derive such \bgraphs from precedence relations as in Definition~\ref{def:precx}: their richer structure makes \bgraphs more expressive---they convey strictly more information---than digraphs.
In contrast, we can easily transform a \bgraph into a precedence relation by splitting \barcs into single-tailed arcs in the obvious way.
Deriving precedence relations from more expressive \bgraphs therefore constitutes a correct way of obtaining strict total orders that satisfy the \textbf{require}ments of Algorithm~\ref{algo:commandification}; doing so just eliminates irrelevant information.

Thus, we propose the following.
Instead of formalizing dependencies among data literals in a set~$\symlitfun(\varphi) \cup \set{\bigstar}$ directly as a precedence relation, we first formalize those dependencies as a \bgraph.
If the resulting \bgraph defines a~$\bigstar$-arborescence, we can directly extract a cycle-free precedence relation~$\precstrict$.
Otherwise, we compute a~$\bigstar$-arborescence of the resulting \bgraph and extract a cycle-free precedence relation~$\precstrict$ afterward.
Either way,~$\precstrict$ denotes a strict partial order whose linearization satisfies the \textbf{require}ments in Algorithm~\ref{algo:commandification}.

\begin{figure}
	\centering

	\renewcommand{\WIDTH}{\linewidth}
	\begin{framed}
		\begin{tabular}{@{}c@{}c@{}}
			\multicolumn{2}{@{}c@{}}{%
				\begin{minipage}{\WIDTH}
					\noindent%
					\begin{equation}
						\label{rule:bprec+free}
						\dfracx[l]{%
							\ell \in \symlitfun(\varphi)
						\\	\LAND \variablfun(\ell) = \set{x_1 \, \ldots \, x_k}
						\\	\LAND x_1 \termeq t_1 \, \ldots \, x_k \termeq t_k \in \symlitfun(\varphi) \cup \setb{\hat x \termeq \hat x}{\hat x \in X}
						}{%
							\set{x_1 \termeq t_1 \, \ldots \, x_k \termeq t_k} \bprec \ell
						}
					\end{equation}
				\end{minipage}
			}
		\\	
		\\	\multicolumn{2}{@{}c@{}}{%
				\begin{minipage}{\WIDTH}
					\noindent%
					\begin{equation}
						\label{rule:bprec+freet}
						\dfracx[l]{%
							x \termeq t \in \symlitfun(\varphi)
						\\	\LAND \variablfun(t) = \set{x_1 \, \ldots \, x_k}
						\\	\LAND x_1 \termeq t_1 \, \ldots \, x_k \termeq t_k \in \symlitfun(\varphi) \cup \setb{\hat x \termeq \hat x}{\hat x \in X}
						}{%
							\set{x_1 \termeq t_1 \, \ldots \, x_k \termeq t_k} \bprec x \termeq t
						}
					\end{equation}
				\end{minipage}
			}
		\\
		\\	\multicolumn{2}{@{}c@{}}{
				\begin{minipage}{\linewidth}
					\noindent%
					\begin{equation}
						\label{rule:bprec+x}
						\dfracx{%
							x \in X
						}{%
							\bigstar \bprec x \termeq x
						}
					\end{equation}
				\end{minipage}
			}
		\end{tabular}
	\end{framed}
	
	\caption{Addendum to Definition~\ref{def:bprec}}
	\label{fig:def+bprec}
\end{figure}

\begin{defi}
	[\textsc{b}-precedence]
	\label{def:bprec}
	${\bprec} : \dcuniv \times \pset{\xuniv} \rightarrow \pset{(\pset{\dcuniv} \cup \set{\bigstar}) \times \dcuniv}$ denotes the function defined by the following equation:
	$$
	{\bprec}(\varphi \, X) = {\bprec}
	$$
	where~$\bprec$ denotes the smallest relation induced by the rules in Figure~\ref{fig:def+bprec}.
\end{defi}

We usually write~$\bprec_\varphi^X$ instead of~${\bprec}(\varphi \, X)$ and use~$\bprec_\varphi^X$ as an infix relation.
Rule~\ref{rule:bprec+free} generalizes Rule~\ref{rule:prec+free} in Definition~\ref{def:prec}, by joining sets of dependencies of a data literal in a single \barc.
Rule~\ref{rule:bprec+freet} states that~$x \termeq t$ does not necessarily depend on~$x$---as implied by Rule~\ref{rule:bprec+free}---but only on the free variables in~$t$ (i.e., we can derive a value for~$x$ from values of the data variables in~$t$).
Note that through Rules~\ref{rule:bprec+free} and~\ref{rule:bprec+freet}, we extend the previous domain~$\symlitfun(\varphi) \cup \set{\bigstar}$ with \emph{semantically insignificant} data equalities of the form~$x \termeq x$, each of which we relate to~$\bigstar$ with Rule~\ref{rule:bprec+x}.
We do this only for the technical convenience of treating both uncontrollable data variables in~$X$ (which may have no data equalities in~$\symlitfun(\varphi)$) and the other variables (which must have \editt{data equalities}) in a uniform way.
For instance, Figure~\ref{fig:bprec} shows the \bgraph for data constraint~$\phieg$.

Generally, in a \bgraph, data literals can have multiple incoming \barcs, which represents a disjunction of conjunctions of dependencies.
Importantly, as long as Algorithm~\ref{algo:commandification} respects the dependencies represented by \emph{one} incoming \barc, the other incoming \barcs do not matter.
An arborescence, which contains one incoming \barc for every data literal, therefore preserves enough dependencies.
Shortly, Theorem~\ref{thm:algo+commandification+requir} makes this more precise.

We can straightforwardly compute an arborescence of a \bgraph
$$
\tpl{\symlitfun(\varphi) \cup \set{\bigstar} \cup \setb{x \termeq x}{x \in X} \, \bprec_\varphi^X}
$$
with an exploration algorithm reminiscent of breadth-first search.
First, let~${\vartriangleleft} \subseteq {\bprec_\varphi^X}$ denote the aborescence under computation, and let~$L_\text{done} \subseteq \symlitfun(\varphi) \cup \set{\bigstar} \cup \setb{x \termeq x}{x \in X}$ denote the set of vertices (i.e., data literals) already explored; initially,~${\vartriangleleft} = \emptyset$ and~$L_\text{done} = \set{\bigstar}$.
Now, given some~$L_\text{done}$, compute a set of vertices~$L_\text{next}$ connected only to vertices in~$L_\text{done}$ by a \barc in~$\bprec_\varphi^X$.
Then, for every vertex in~$L_\text{next}$, add an incoming \barc to~$\vartriangleleft$.%
\footnote{%
	\label{foot:complex}%
	If a vertex~$\ell$ in~$L_\text{next}$ has multiple incoming \barcs, the choice among them matters not: the choice remains local, because every \barc has only one head (i.e., adding an~$\ell$-headed \barc to~$\vartriangleleft$ cannot cause another vertex to get multiple incoming \barcs, which would invalidate the arborescence).
	General hypergraphs, whose hyperarcs can have multiple heads, violate this property (i.e., the choice of which hyperarc to add becomes global instead of local).\
	As a result, and in stark constrast to \bgraphs, one cannot compute arborescences of general hypergraphs---an \textsc{np}-complete problem~\cite{Woe92}---in polynomial time (if~$\textsc{p} \neq \textsc{np}$).
}
Afterward, add~$L_\text{next}$ to~$L_\text{done}$.
Repeat this process until~$L_\text{next}$ becomes empty.
Once that happens, either~$\vartriangleleft$ contains an arborescence (if~$L_\text{done} = L$) or no arborescence exists.
This computation runs in linear time, in the size of the \bgraph.
See also Footnote~\ref{foot:complex}.
Henceforth, let~$\bprecarbor_\varphi^X$ denote the final arborescence so computed; if no arborescence exists, we stipulate~$\bprecarbor_\varphi^X = \emptyset$.

\begin{figure}
	\centering
	
	\renewcommand{\WIDTH}{.5\linewidth}
	\begin{framed}
		\begin{tabular}{@{}c@{}c@{}}
			\begin{minipage}{\WIDTH}
				\noindent%
				\begin{equation}
					\label{rule:precstrict+free}
					\hphantom{x} \dfracx[l]{%
						\ell_1 \in \symlitfun(\varphi) \cap L
					\\	\LAND L \bprecarbor_\varphi^X \ell_2
					}{%
						\ell_1 \precstrict \ell_2
					} \hphantom{x}
				\end{equation}
			\end{minipage}
		&	\begin{minipage}{\WIDTH}
				\noindent%
				\begin{equation}
					\label{rule:precstrict+conv}
					\dfracx[l]{%
						x \termeq t \, \ell \in \symlitfun(\varphi)
					\\	\LAND \SCOPE{\ell \neq x' \termeq t' \FORALL x' \, t'}
					}{%
						x \termeq t \precstrict \ell
					}
				\end{equation}
			\end{minipage}
		\\
		\\	\multicolumn{2}{@{}c@{}}{%
				\begin{minipage}{\linewidth}
					\noindent%
					\begin{equation}
						\label{rule:precstrict+trans}
						 \dfrac{%
							\ell_1 \precstrict \ell_2 \AND \ell_2 \precstrict \ell_3 \AND \ell_2 \notin \set{\ell_1 \, \ell_3}
						}{%
							\ell_1 \precstrict \ell_3
						}
					\end{equation}
				\end{minipage}
			}
		\end{tabular}
	\end{framed}
	
	\caption{Addendum to Definition~\ref{def:precstrict}}
	\label{fig:def+precstrict}
\end{figure}

\begin{defi}
	[precedence \III]
	\label{def:precstrict}
	${\precstrict} : \dcuniv \times \pset{\xuniv} \rightarrow \dcuniv \times \dcuniv$ denotes the function defined by the following equation:
	$$
	{\precstrict}(\varphi \, X) = {\precstrict}
	$$
	where~$\precstrict$ denotes the smallest relation induced by the rules in Figure~\ref{fig:def+precstrict}.
\end{defi}

We usually write~$\precstrict_\varphi^X$ instead of~${\precstrict}(\varphi \, X)$.
Rules~\ref{rule:precstrict+conv} and~\ref{rule:precstrict+trans} have the same premise\slash con\-se\-quence as Rules~\ref{rule:prec+conv} and~\ref{rule:prec+trans}; Rule~\ref{rule:precstrict+free} straightforwardly splits \barcs into single-tailed arcs.
For instance, the bold arcs in Figure~\ref{fig:prec} represent a fragment of the precedence relation so derived from the arborescence in Figure~\ref{fig:bprec}.

For every~$\precstrict_\varphi^X$ induced from a nonempty~$\bigstar$-arborescence (i.e.,~${\bprecarbor_\varphi^X} \neq \emptyset$), let~$\prectotal_\varphi^X$ denote its linearization.
The following theorem states that this linearization satisfies the \textbf{require}ments of Algorithm~\ref{algo:commandification}.

\begin{thm}
	\label{thm:algo+commandification+requir}
	$${\bprecarbor_\varphi^X} \neq \emptyset \IMPLIES \SCOPE{\tpl{\varphi \, X \, \prectotal_\varphi^X} \text{ satisfies the \textbf{require}ments of Algorithm~\ref{algo:commandification}}}$$
\end{thm}\medskip

\noindent If the \bgraph~$\tpl{\symlitfun(\varphi) \cup \set{\bigstar} \cup \setb{x \termeq x}{x \in X} \, \bprec_\varphi^X}$ neither defines nor contains a~$\bigstar$-arborescence, no \bgraph equivalent of a path~\cite{AFF01} exists from~$\bigstar$ to at least one vertex~$\ell$.
In that case, the other vertices fail to resolve at least one of~$\ell$'s dependencies.
This occurs, for instance, when~$\ell$ depends on~$x$, but the \bgraph contains no~$x \termeq t$ vertex.
As another example, consider a recursive data equality~$x \termeq t$ with~$x \in \variablfun(t)$: unless another data equality~$x \termeq t'$ with~$t \neq t'$ exists, \editt{every} incoming \editt{\barc in its \bgraph loops onto} itself.
Consequently, no arborescence exists.
In practice, such cases inherently require constraint solving techniques with backtracking to find a value for~$x$.
Nonexistence of a~$\bigstar$-arborescence thus signals a hard limit to the applicability of Algorithm~\ref{algo:commandification} (although mixed techniques of translating some parts of a data constraint to a data command at compile-time and leaving other parts to a constraint solver at run-time seem worthwhile to explore; we leave \editt{this possibility} for future work).
Thus, the set of data constraints to which we can apply Algorithm~\ref{algo:commandification} contains those (i) whose \bgraph has a~$\bigstar$-arborescence, which guarantees linearizability of the induced precedence, and (ii) that satisfy also the rest of the \textbf{require}ments in Algorithm~\ref{algo:commandification}.

\subsection*{Commandify}

To introduce data commands in \CAs, we introduce \emph{commandify} as a unary operation on \CAs.
First, because we want to avoid ad-hoc modifications to Definitions~\ref{def:dcuniv} and~\ref{def:automuniv} (of data constraints and \CAs), we present an encoding of data commands as data relations.
In the following definition, let~$\varphi$ denote a data constraint in a \CA, let~$X$ denote the set of uncontrollable data variables in~$\varphi$, and let~$x_1 \, \ldots \, x_k$ denote the free data variables in~$\varphi$, ordered by~$\termorder$.
Then, data relation~$R$, which encodes the commandification~$\pi$ of~$\varphi$, holds true of a data tuple~$\tpl{d_1 \, \ldots \, d_k}$ iff execution of~$\pi$ on an initial data state (over the variables in~$X$) successfully terminates on a data state~$\sigma$ that maps every~$x_i$ to~$d_i$.

\begin{defi}
	[data commands as data relations]
	\label{def:commfun}
	$\commfun : \dcuniv \times \pset{\xuniv} \rightarrow \dcuniv$ denotes the function defined by the following equation:
	$$
	\commfun(\varphi \, X) = \left \{ \begin{array}{@{} l @{} l @{}}
		R(x_1 \, \ldots \, x_k)	& \IF \SCOPEX[l]{%
									\freefun(\varphi) = \set{x_1 \, \ldots \, x_k}
								\\	\LAND x_1 \termorder \cdots \termorder x_k
								\\	\LAND {\bprecarbor_\varphi^X} \neq \emptyset
								\\	\LAND X \subseteq \freefun(\varphi)
								}
	\\	\UP
	\\	\varphi					& \OTHERWISE
	\end{array} \right .
	$$
	where~$R$ denotes the smallest relation induced by the following rule:
	\begin{equation}
		\label{rule:commfun}
		\dfracx[l]{%
				\pi = \textsc{Algorithm\ref{algo:commandification}}(\varphi \, X \, \precstrict_\varphi^X)
			\\	\LAND \sigma \in \finalfun(\pi \, \dcchar{\bigwedge \setb{x \termeq x}{x \in X}})
			\\	\LAND \sigma(x_1) \, \ldots \, \sigma(x_k) \in \duniv
		}{%
			\tpl{\sigma(x_1) \, \ldots \, \sigma(x_k)} \in R
		}
	\end{equation}
\end{defi}

\noindent
Note that~$\sigma$ in Rule~\ref{rule:commfun} may map also data variables outside~$\freefun(\varphi)$.
This happens, for instance, with data constraints with existential quantifiers.
The data commands for such data constraints explicitly assign values to quantified data variables, even though those variables do not qualify as free.
Because~$\set{x_1 \mapsto d_1 \, \ldots \, x_k \mapsto d_k}$ contains the free data variables in~$\varphi$, however, the additional data variables mapped by~$\sigma$ cannot affect the truth of~$\varphi$ (by monotonicity of entailment).

We define commandification in \CAs in terms of~$\commfun$.

\begin{defi}
	[commandify]
	\label{def:comm}
	$\comm{\cdot} : \automuniv \rightarrow \automuniv$ denotes the function defined by the following equation:
	$$
	\comm{\tpl{Q \, \tpl{P^\textall \, P^\textin \, P^\textout} \, M \, \longrightarrow \, q^0}} = \tpl{Q \, \tpl{P^\textall \, P^\textin \, P^\textout} \, M \, \commr{\longrightarrow} \, q^0 \, \mu^0}
	$$
	where~$\commr{\longrightarrow}$ denotes the smallest relation induced by the following rules:
	\begin{center}
		\begin{tabular}{@{}c@{}}
			\begin{minipage}{\linewidth}
				\noindent%
				\begin{equation}
					\dfracx[l]{%
						q \xrightarrow{P,\varphi} q' \AND X^\textinit = P^\textin \cup \pre{M}
					}{%
						q \commr{\xrightarrow{P,\commfun(\varphi,\freefun(\varphi)\cap X^\textinit)}} q'
					}
				\end{equation}
			\end{minipage}
		\end{tabular}
	\end{center}
\end{defi}

\subsection*{Correctness and Effectiveness}

We conclude this section by establishing the correctness and effectiveness of commandify.
We consider commandify correct if it yields a behaviorally con\-gru\-ent \CA to the original one.
Before formulating this as a theorem, the following lemma first states the equivalence of a data constraint and its commandification.

\begin{lem}
	\label{lemma:dcequiv+commfun}
	$\varphi \dcequiv \commfun(\varphi \, X)$
\end{lem}

From Proposition~\ref{prop:001} and Lemma~\ref{lemma:dcequiv+commfun}, we conclude the following correctness theorem.

\begin{thm}
	\label{thm:congr+comm}
	$\bfa \congr \comm{\bfa}$
\end{thm}

We consider commandify effective if, after commandifying a \CA~$\bfa$, every data constraint in \editt{the resulting} \CA either encodes a data command as in Definition~\ref{def:commfun} or has no data variables in it (in which case a compiler can statically check that data constraint).
Generally, however, such unconditional effectiveness does not hold true.
After all, if the \bgraph for a data constraint~$\varphi$ in~$\bfa$ has no~$\bigstar$-arborescence, we have no strict precedence relation to run Algorithm~\ref{algo:commandification} with.
In that case,~$\commfun(\varphi \, X) = \varphi$, and consequently, commandify does not have its intended effect.
Fortunately, commandify does satisfy a weaker---but useful---form of effectiveness.
To formulate this as a theorem, we first define a relation that holds true of \emph{arborescent} \CAs.
We consider a \CA arborescent if the \bgraph for each of its data constraints has a~$\bigstar$-arborescence.

\begin{defi}
	[arborescentness]
	\label{def:arbor}
	$\arbor \subseteq \automuniv$ denotes the smallest relation induced by the following rule:
	\begin{equation}
		\label{eqn:arbor}
		\dfrac{%
			\SCOPE{\varphi \in \dcfun(\bfa) \IMPLIES {\bprecarbor_\varphi^X} \neq \emptyset} \FORALL \varphi
		}{%
			\arbor \bfa
		}
	\end{equation}
\end{defi}\medskip

\noindent
The following theorem states the effectiveness of commandify, conditional on arborescentness: after commandifying an arborescent \CA~$\bfa$, every data constraint in \editt{the resulting} \CA encodes a data command as a data relation (as in Definition~\ref{def:commfun}).
Let~$R$ range over the set of data relations defined in Definition~\ref{def:commfun} of~$\commfun$.

\begin{thm}
	\label{thm:comm+effect}
	$\arbor{\bfa} \IMPLIES \dcfun(\comm{\bfa}) \subseteq \setb{R(x_1 \, \ldots \, x_k)}{\TRUE}$
\end{thm}

\subsection*{Discussion}

The constraint programming community has already observed that, for constraint solving, ``if domain specific methods are available they should be applied \emph{instead} [sic] of the general methods''~\cite{Apt09a}.
Commandification pushes this piece of conventional wisdom to an extreme: essentially, every data command generated for a data constraint~$\varphi$ by Algorithm~\ref{algo:commandification} constitutes a \editt{small}, dedicated constraint solver capable of solving only~$\varphi$.
Nevertheless, execution of data commands bears similarities with \emph{constraint propagation} techniques, in particular with \emph{forward checking}~\cite{BMFL02}.
Generally, constraint propagation aims to reduce the search space of a constraint satisfaction problem by transforming it into an equivalent ``simpler'' one, where variables have smaller domains, or where constraints refer to fewer variables.
With forward checking, whenever a variable~$x$ gets a value~$d$, a constraint solver removes values from the domains of all subsequent variables that, given~$d$, violate a constraint. 
In the case of an equality~$x = x'$, for instance, forward checking reduces the domain of~$x'$ to the singleton~$\set{d}$ after an assignment of~$d$ to~$x$.
Commandification implicitly uses that same property of equality, but instead of explicitly representing the domain of a variable and the reduction of this domain to a singleton at run-time, commandification already turns the equality into an assignment at compile-time.

Commandification may also remind one of classical \emph{Gaussian eliminination} for solving systems of linear equations over the reals~\cite{Apt09b}: there too, one orders variables and substitutes values/expressions for variables in other expressions.
Data constraints, however, have a significantly different structure from real numbers, which makes solving data constraints directly via Gaussian elimination at least not obvious.

Before we did the work presented in this paper, Clarke et al. already worked on purely constraint-based implementations of protocols~\cite{CPLA11}.
Essentially, Clarke et al. specify not only the transition labels of an automaton as boolean constraints but also its state space and transition relation.
In recent work, Proen\c{c}a and Clarke developed a variant of compile-time \emph{predicate abstraction} to improve performance~\cite{PC13a}.
They also used this technique to allow a form of interaction between a constraint solver and its environment during constraint solving~\cite{PC13b}.
The work of Proen\c{c}a and Clarke resembles our work in the sense that we all try to ``simplify'' constraints at compile-time.
We see also differences, though: (i) commandification fully avoids constraint solving and (ii) we adopted a richer language of data constraints in this paper.
For instance, Proen\c{c}a and Clarke have only unary functions in
their language, which would have \editt{avoided} our need for
\bgraphs.

%
\section{Experiments}
\label{sect:exper}
\subsection*{Setup}

We implemented our two optimization techniques as extensions to our existing \CA-to-Java compiler, a plug-in for the Eclipse \IDE[].
This plug-in is an integrated part of a larger toolset, which also consists of an editor that supports the graphical syntax for \CAs presented in Section~\ref{sect:prel}, through a drag-and-drop interface.
To evaluate the impact of our optimization techniques in practice, then, we performed a number of experiments with their implementation, the results of which we present in this section.

We divided our experiments into two categories.
The first category consists of experiments involving \compilergenerated coordinator threads in isolation.
These experiments are ``pure'' in the sense that we measure only the performance of the \compilergenerated code, without ``polluting'' these measurements with delays caused by process threads.
The second category consists of experiments involving \compilergenerated coordinator threads \emph{in the context of full programs}.
These experiments allow us to observe the impact of our optimization techniques on the performance of full programs.

We ran each of our experiments five times on a machine with~$24$ cores (two Intel E5-2690V3 processors in two sockets), without Hyper-Threading and without Turbo Boost (i.e., with a static clock frequency), and averaged our measurements afterward.

\subsection*{Category I}

To study the performance of \compilergenerated coordinator threads in \edit{isolation}, we selected seven sets of \CAs for experimentation, whose elements differ in the value of~$k \in \set{1 \, 2 \, 3 \, 4 \, 6 \, 8 \, 12 \, 16 \, 24 \, 32 \, 48 \, 64}$: \reo{Sync}$_k$, \reo{Fifo}$_k$, \reo{OddFib}$_k$, \reo{Merg}$_k$, \reo{LateAsyncMerg}$_k$, \reo{EarlyAsyncMerg}$_k$, and \reo{Rout}$_k$.
In total, thus, we generated code for~$96$ \CAs, yielding~$96$ experiments.
\edit{Application of our optimization techniques did not add any measurable compilation overhead.}
Each of these \CAs, except the \reo{Merg}$_k$ \CAs, is the~$k$-parametric generalization of a \CA denoted by a digraph in Figure~\ref{fig:reo+oddfibonacci+2}; every \reo{Merg}$_k$ \CA is the~$k$-parametric generalization of \reo{Merg2} in Figure~\ref{fig:autom+core}.
For \reo{Sync}$_k$\slash \reo{Fifo}$_k$, parameter~$k$ controls the number of \reo{Sync}s\slash\reo{Fifo}s in the chain.
For \reo{Merg}$_k$, \reo{LateAsyncMerg}$_k$, and \reo{EarlyAsyncMerg}$_k$, parameter~$k$ controls the number of producers.
For \reo{OddFib}$_k$ and \reo{Rout}$_k$, parameter~$k$ controls the number of consumers.
See Section~\ref{sect:prel} for a brief description of the behavior of these \CAs for~$k = 2$.

In each run of an experiment, we measured the number of completed transitions in four minutes after warming up the Java virtual machine for thirty seconds.
To measure the performance of only the compiler-generated code, we used ``empty'' producers and consumers, which essentially execute \code{while\:(true)\:put(...)} and \code{while\:(true)\:get(...)}.

\begin{figure}[t]
	\subfloatvspace
	\renewcommand{\reosize}{\scriptsize}
	\captionsetup[subfloat]{labelformat=empty}
	\hfil%
	\subfloat[Legend]{\includegraphics{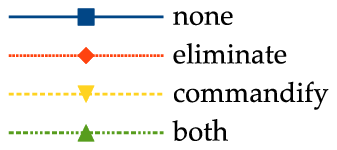}}%
	\hfil%
	\subfloat[\reo{Sync}]{\includegraphics{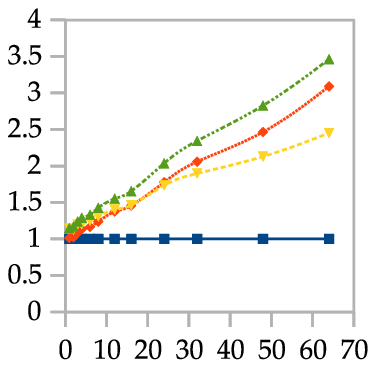}}%
	\hfil%
	\subfloat[\reo{Fifo}]{\includegraphics{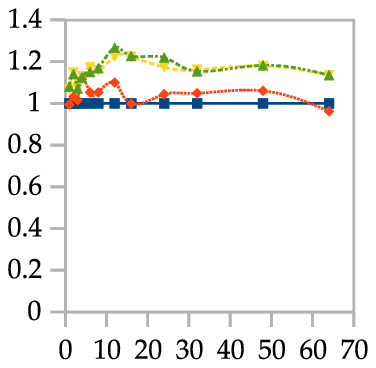}}%
	\hfil%
	\subfloat[\reo{OddFib}]{\includegraphics{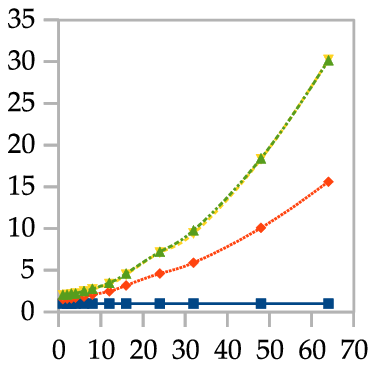}}%
	\hfil%
	
	\hfil%
	\subfloat[\reo{Merg}]{\includegraphics{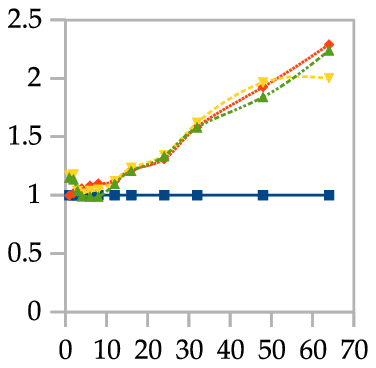}}%
	\hfil%
	\subfloat[\reo{LateAsyncMerg}]{\includegraphics{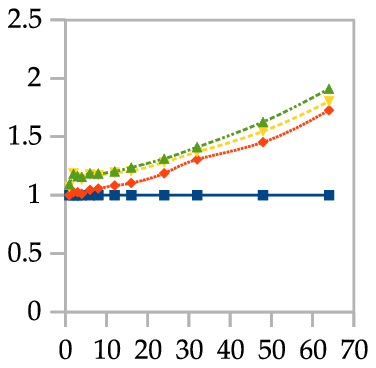}}%
	\hfil%
	\subfloat[\reo{EarlyAsyncMerg}]{\includegraphics{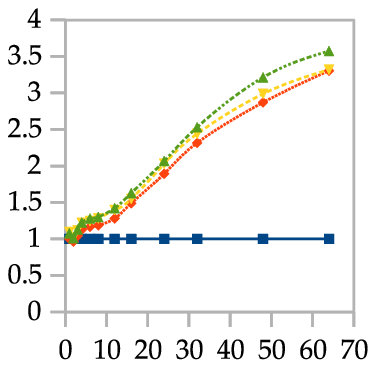}}%
	\hfil%
	\subfloat[\reo{Rout}]{\includegraphics{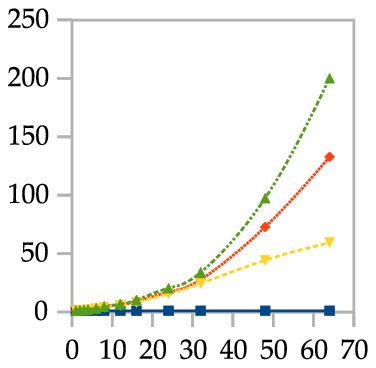}}%
	\hfil%
	
	\caption{Experimental results for seven sets of \CAs in isolation: speedups, on the y-axis, of compiler-generated code optimized with eliminate, commandify, or both, relative to unoptimized code, as a function of the number of processes, on the x-axis}
	\label{fig:exper-prot}
\end{figure}

Figure~\ref{fig:exper-prot} shows our experimental results.
The figure shows that, individually, our two optimization techniques are already very effective.
When we apply both optimization techniques simultaneously, in many cases (\reo{Sync}$_k$, \reo{LateAsyncMerg}$_k$, \reo{EarlyAsyncMerg}$_k$, and \reo{Rout}$_k$), performance is further improved, but the improvement is not the sum of the individual improvements.
The reason is that after applying one of the techniques, there is ``less room'' for the other technique \editt{to make} further improvement: there is only so much that can be optimized \editt{in} checking data constraints, and each of our two techniques individually seems to already \editt{make} a significant step toward an optimum.
Still, as Figure~\ref{fig:exper-prot} shows, it is useful to apply both techniques, especially since they do not appear to negatively influence each other.

\subsection*{Category II}

To study the performance of \compilergenerated coordinator threads in the context of full programs, we adapted the \emph{\NAS[] Parallel Benchmarks} \NPB[]~\cite{BBB+91}, a popular suite to evaluate parallel performance with.
The \NPB[] suite specifies eight benchmarks---five computational kernels and three realistic applications---derived from computational fluid dynamics programs; for each of these benchmarks, to standardize comparisons, the \NPB[] suite specifies four classes of problem sizes (\classw, \classa, \classb, \classc).

We compared the Java reference implementation of \NPB[] with a \CA-based implementation.
The Java reference implementation, developed by Frumkin et al.~\cite{FSJY03}, contains a Java program for seven of \NPB[]'s eight benchmarks; one kernel benchmark is missing.
Each of these programs consists of a \emph{master} process and a number of \emph{worker} processes.
The master and its workers interact with each other under a classical master\slash workers protocol (i.e., the master distributes work among its workers; the workers inform their master once their work is done).
Frumkin et al. programmed this protocol using monitors.

We took the Java reference implementation of \NPB[] as the basis for our \CA-based implementation.
First, we removed all instances of the master\slash workers protocol from the seven programs.
Then, we added ports and \code{put}\slash \code{get}.
Separately, we drew the master\slash workers protocol in our graphical syntax for \CAs.
Subsequently, we compiled our specification for~$k \in \set{2 , 4 , 8 , 16 , 32 , 64}$ workers (unless a combination of benchmark+class supported only fewer workers), and let our compiler automatically integrate the \handwritten code (for masters\slash workers) with its own \compilergenerated code.
\edit{Application of our optimization techniques did not add any measurable compilation overhead.}

\begin{figure}[t]
	\subfloatvspace
	\renewcommand{\reosize}{\scriptsize}
	\captionsetup[subfloat]{labelformat=empty}
	\hfill%
	\subfloat[Legend]{\includegraphics{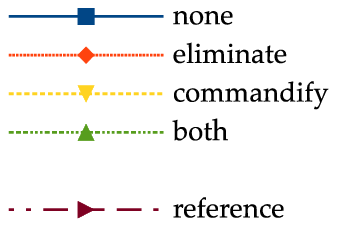}}%
	\hfill%
	\subfloat[{\NPBBT[]}: \classw]{\includegraphics{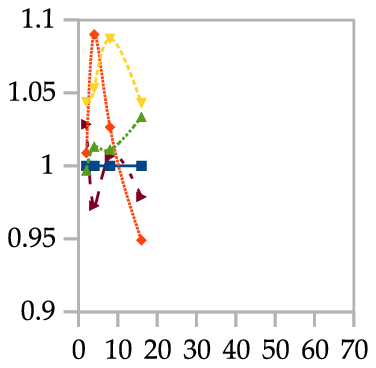}}%
	\hfil%
	\subfloat[{\NPBBT[]}: \classa]{\includegraphics{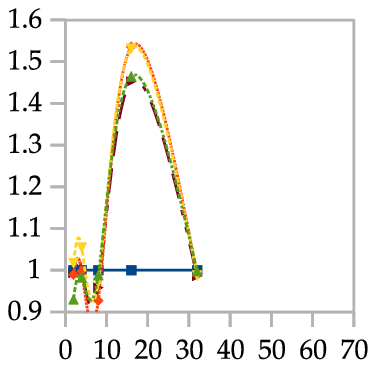}}%
	\hfil%
	
	\hfil%
	\subfloat[{\NPBLU[]}: \classw]{\includegraphics{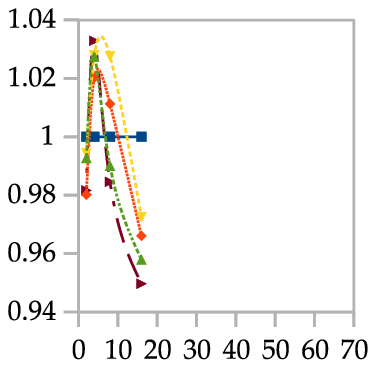}}%
	\hfil%
	\subfloat[{\NPBLU[]}: \classa]{\includegraphics{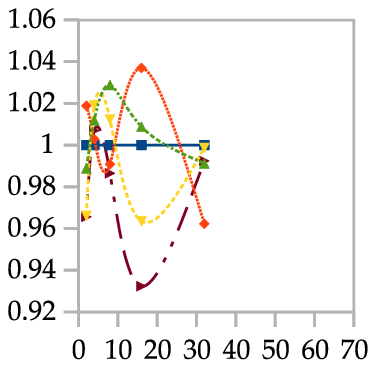}}%
	\hfil%
	\subfloat[{\NPBSP[]}: \classw]{\includegraphics{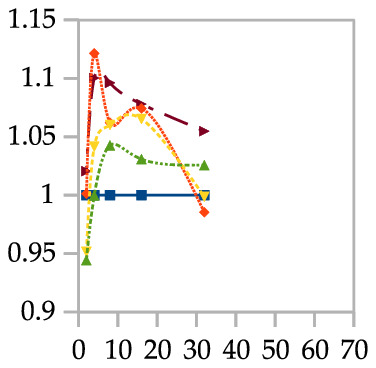}}%
	\hfil%
	\subfloat[{\NPBSP[]}: \classa]{\includegraphics{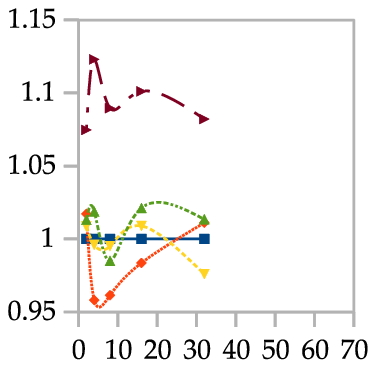}}%
	\hfil%
	
	\caption{Experimental results for three \NPB[] applications: speedups (y-axis) of compiler-generated code optimized with eliminate, commandify, or both, and of reference code by Frumkin et al., relative to unoptimized compiler-generated code, as a function of the number of processes (x-axis)}
	\label{fig:exper-npb-appl}
\end{figure}

\begin{figure}[p]
	\subfloatvspace
	\renewcommand{\reosize}{\scriptsize}
	\captionsetup[subfloat]{labelformat=empty}
	\hfil%
	\subfloat[{\NPBCG[]}: \classw]{\includegraphics{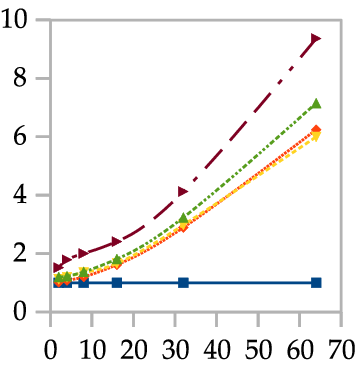}}%
	\hfil%
	\subfloat[{\NPBCG[]}: \classa]{\includegraphics{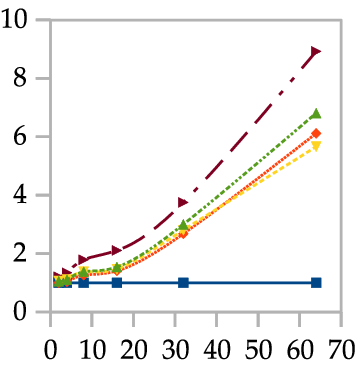}}%
	\hfil%
	\subfloat[{\NPBCG[]}: \classb]{\includegraphics{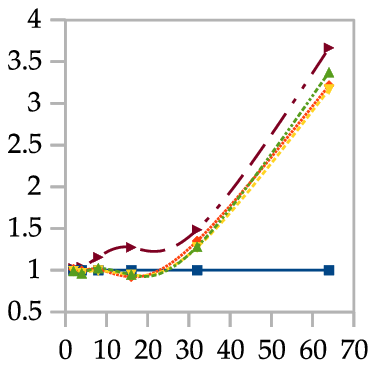}}%
	\hfil%
	\subfloat[{\NPBCG[]}: \classc]{\includegraphics{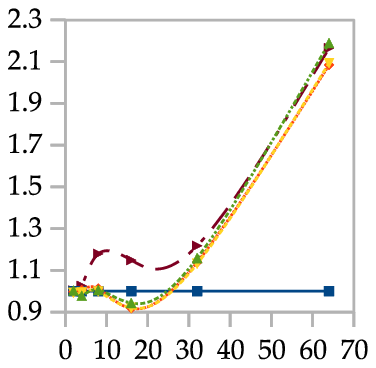}}%
	\hfil%
	
	\hfil%
	\subfloat[{\NPBFT[]}: \classw]{\includegraphics{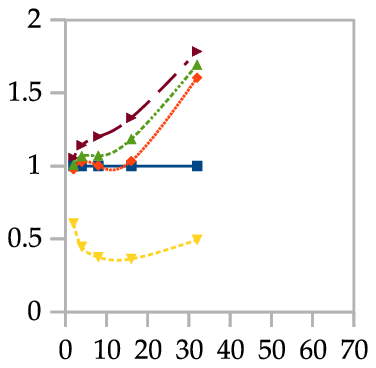}}%
	\hfil%
	\subfloat[{\NPBFT[]}: \classa]{\includegraphics{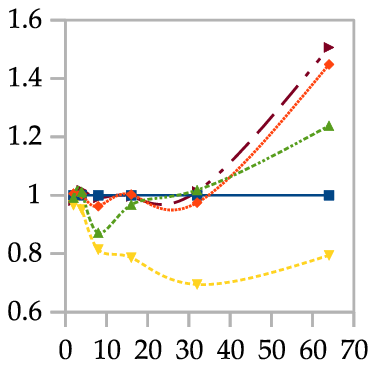}}%
	\hfil%
	\subfloat[{\NPBFT[]}: \classb]{\includegraphics{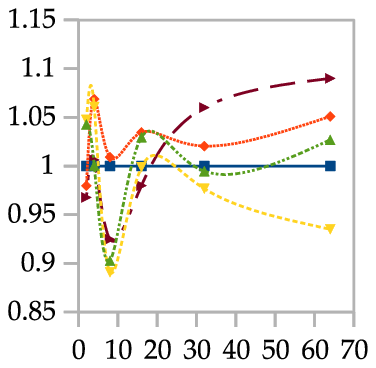}}%
	\hfil%
	\subfloat[{\NPBFT[]}: \classc]{\includegraphics{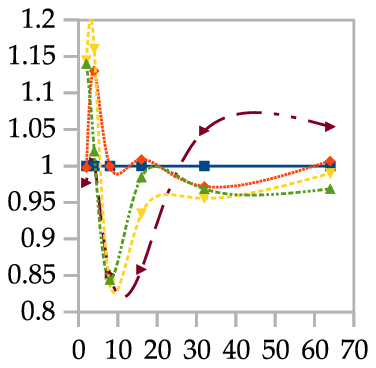}}%
	\hfil%
	
	\hfil%
	\subfloat[{\NPBIS[]}: \classw]{\includegraphics{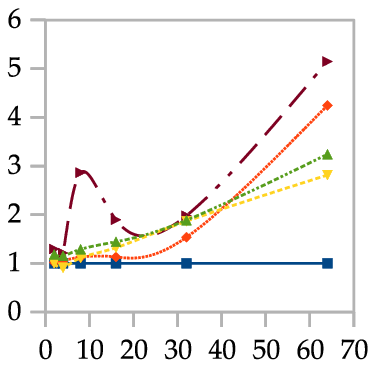}}%
	\hfil%
	\subfloat[{\NPBIS[]}: \classa]{\includegraphics{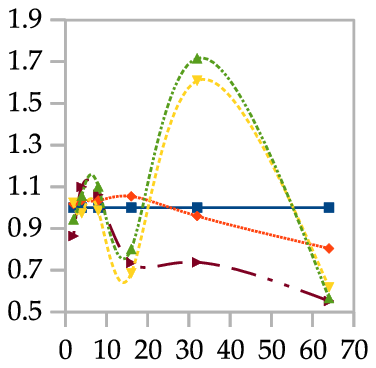}}%
	\hfil%
	\subfloat[{\NPBIS[]}: \classb]{\includegraphics{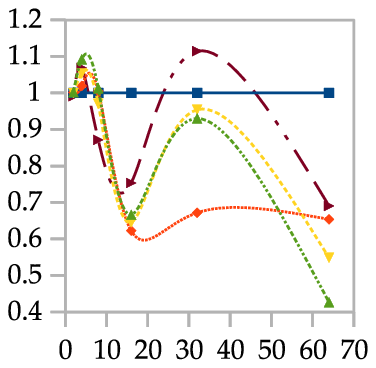}}%
	\hfil%
	\subfloat[{\NPBIS[]}: \classc]{\includegraphics{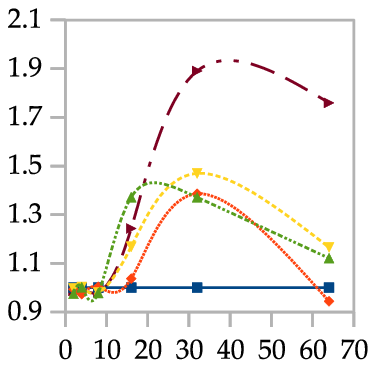}}%
	\hfil%
	
	\hfil%
	\subfloat[{\NPBMG[]}: \classw]{\includegraphics{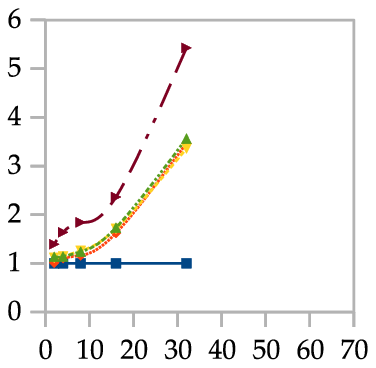}}%
	\hfil%
	\subfloat[{\NPBMG[]}: \classa]{\includegraphics{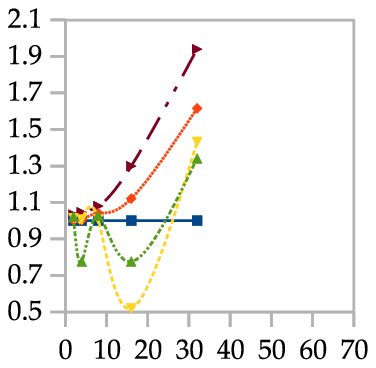}}%
	\hfil%
	\subfloat[{\NPBMG[]}: \classb]{\includegraphics{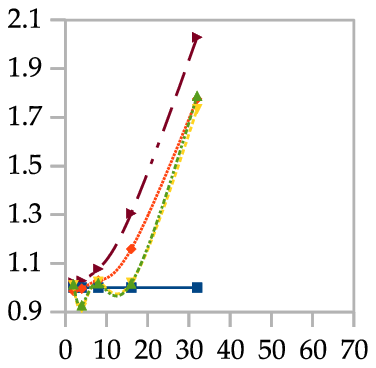}}%
	\hfil%
	\subfloat[{\NPBMG[]}: \classc]{\includegraphics{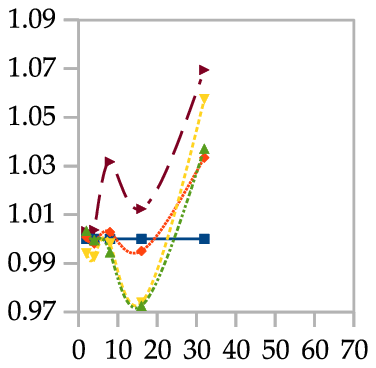}}%
	\hfil%
	
	\caption{Experimental results for four \NPB[] kernels: speedups, on the y-axis, of compiler-generated code optimized with eliminate, commandify, or both, and of reference code by Frumkin et al., relative to unoptimized compiler-generated code, as a function of the number of processes, on the x-axis. See Figure~\ref{fig:exper-npb-appl} for a legend.}
	\label{fig:exper-npb-kern}
\end{figure}

Figures~\ref{fig:exper-npb-appl} and~\ref{fig:exper-npb-kern} show our experimental results.
These results, in contrast to the results in Figure~\ref{fig:exper-prot}, look messy and are hard to derive a meaningful conclusion from: in some cases, using both optimizations results in the best performance, but in other cases, using only one of the optimizations results in the best performance, and in yet a few other cases, using \emph{no} optimization actually results in the best performance.

The reason for these results, so we found out, has to do with hardware cache performance: it turns out that the memory footprint of our \compilergenerated code seriously impacts numbers of cache misses, a phenomenon that did not yet manifest when we ran our \compilergenerated code in isolation.
As we have not yet optimized \compilergenerated code for memory usage, a reasonable assumption is that code with a large memory footprint results in more cache misses.
However, things are even more subtle than that: due to the way the Java virtual machine allocates memory, so we found out, a \emph{larger} memory footprint may in fact result in \emph{fewer} cache misses.
\editt{We admit that we do not yet understand the impact of the memory footprint of our \compilergenerated code on the execution-time performance of the code sufficiently well enough to appropriately account for this impact in our optimization schemes.
This investigation constitutes an important piece of our future work.
We consider the revelation of this underdeveloped aspect of our
compilation technology as a significant contribution of this paper.}

%
\section{Conclusion}
\label{sect:concl}

We presented, and established the correctness of, two techniques to optimize the performance of checking data constraints.
The first technique, called ``eliminate'' and formalized as operation~$\subtry$, reduces the size of data constraints at compile-time, to reduce the complexity of constraint solving at run-time.
The second technique, called ``commandify'' and formalized as operation~$\comm{\cdot}$, translates data constraints into \editt{small} pieces of imperative code at compile-time, to replace expensive calls to a general-purpose constraint solver at run-time.
\editt{Finding satisfying assignments for data constraints resembles a game of hide-and-seek, played by our \compilergenerated code at run-time with the aid of a constraint solver.
This game was reasonable when our \CA compilation technology was still in its infancy, but no longer as this technology matures.}

Although the experiments in which we evaluated \compilergenerated code in isolation show that eliminate and commandify indeed have a positive impact on performance, the experiments in which we evaluated \compilergenerated code in the context of full programs remain inconclusive because of seemingly erratic hardware cache behavior.
Here lies an important next research step: we need to better understand the impact of memory footprints of \compilergenerated code.
So far, including in this paper, we have focused our attention exclusively on compilation techniques for optimizing ``algorithmic'' aspects of \compilergenerated code (i.e., minimizing the number of computation steps necessary to, for instance, check data constraints).
Our experimental results in this paper show that we need to start considering memory too.

\edit{Another interesting piece of future work \editt{involves} comparing our compilation technology for constraint automata, including the optimization techniques presented in this paper, with compilation technology for other coordination models and languages.
One interesting candidate is \BIP[].
In recent work~\cite{DJAB15}, we already performed a theoretical study on the relation between (the formal semantics of) Reo and \BIP[].
A natural next step in this line of work \editt{consists of} a practical comparison \editt{of these models} (including not only performance \editt{of their generated code}, but also such software engineering qualities as programmability, maintainability, reusability, and so on).}	
	
	\bibliographystyle{alpha}
	\bibliography{main}
	
	\appendix

%
\section{Proofs}
\label{sect:proofs}
\subsection*{Proof of Lemma~\ref{lemma:dcequiv+exists+existsfun}}
\label{sect:proofs:lemma+dcequiv+exists+existsfun}

If~$p$ has no determinant in~$\varphi$, we have~$\existsfun_p(\varphi) = \exists p . \varphi$, and we are done (because~$\dcequiv$ is reflexive).

Therefore, suppose that~$p$ has a determinant in~$\varphi$, and let~$t$ denote the least such determinant under~$\termorder$ such that~$\existsfun_p(\varphi) = \varphi [t / p]$.
By the grammar of data constraints,~$\varphi$ must be of the form~$\exists p_1 . \cdots \exists p_l . (\allowbreak \ell_1 \wedge \cdots \wedge \ell_k)$.
Thus, we must show~$\exists p . \exists p_1 . \cdots \exists p_l . (\ell_1 \wedge \cdots \wedge \ell_k) \dcequiv (\exists p_1 . \cdots \exists p_l . (\ell_1 \wedge \cdots \wedge \ell_k)) [t / p]$.
To show this, \editt{without loss of generality, we assume~$p, p_1, p_2, \ldots, p_l$ are all distinct (otherwise we can simply eliminate the quantifier of every duplicate variable).
By} the usual definitions of logical equivalence and entailment, we must show that~$\sigma \dcmodels \exists p . \exists p_1 . \cdots \exists p_l . (\ell_1 \wedge \cdots \wedge \ell_k)$ implies~$\sigma \dcmodels (\exists p_1 . \cdots \exists p_l . (\ell_1 \wedge \cdots \wedge \ell_k)) [t / p]$, for all~$\sigma$, and vice versa.

Suppose~$\sigma \dcmodels \exists p . \exists p_1 . \cdots \exists p_l . (\ell_1 \wedge \cdots \wedge \ell_k)$.
By the usual semantics of~$\exists$, this implies~$\sigma \dcmodels (\exists p_1 . \cdots \allowbreak \exists p_l . (\ell_1 \wedge \cdots \wedge \ell_k)) [d / p]$ for some datum~$d$.
\editt{Because}~$p$ is not bound by another~$\exists$, we can expand also the other existential quantifications, and distribute the resulting substitutions over the conjunction, to get~$\sigma \dcmodels \ell_i [d / p] [d_1 / p_1] \cdots [d_l / p_l]$ for every~$\ell_i$.
Now, because~$t$ is a determinant of~$p$, a literal~$t \termeq p$ (or, symmetrically,~$p \termeq t$) must exist among the~$\ell_i$ literals.
So, for that literal, we have~$\sigma \dcmodels t [d / p] [d_1 / p_1] \cdots [d_l / p_l] \termeq d$.
A literal~$t_1 \termeq t_2$ holds under~$\sigma$ iff the evaluation of~$t_1$ equals the evaluation of~$t_2$.
Hence, we know that the evaluation of~$t [d / p] [d_1 / p_1] \cdots [d_l / p_l]$ equals~$d$.
From this, combined with the previous result~$\sigma \dcmodels (\exists p_1 . \cdots \exists p_l . (\ell_1 \wedge \cdots \wedge \ell_k)) [d / p]$, we can establish~$\sigma \dcmodels (\exists p_1 . \cdots \exists p_l . (\ell_1 \wedge \cdots \wedge \ell_k)) [t / p]$.

In the opposite direction, suppose~$\sigma \dcmodels (\exists p_1 . \cdots \exists p_l . (\ell_1 \wedge \cdots \wedge \ell_k)) [t / p]$.
We know that~$\evalfun_\sigma(t) = d$ for some~$d$, as before.
In other words, there exists a~$d$ (namely~$\evalfun_\sigma(t)$) such that~$\sigma \dcmodels (\exists p_1 . \cdots \exists p_l . (\ell_1 \wedge \cdots \wedge \ell_k)) [d / p]$.
By the usual semantics of~$\exists$, this implies~$\sigma \dcmodels \exists p . \exists p_1 . \cdots \exists p_l . (\ell_1 \wedge \cdots \wedge \ell_k)$.

A full, detailed proof appears as the proof of Lemma 16 in~\cite[Appendix D.3]{techreport}.
\qed

\subsection*{Proof of Theorem~\ref{thm:congr+subtrx+subtry}}
\label{sect:proofs:thm+congr+subtrx+subtry}

Follows from Proposition~\ref{prop:001} and Lemma~\ref{lemma:dcequiv+exists+existsfun}.

A full, detailed proof appears as the proof of Theorem 14 in~\cite[Appendix D.3]{techreport}.
\qed

\subsection*{Proof of Theorem~\ref{thm:subtry+effect}}
\label{sect:proofs:thm+subtry+effect}

Reasoning toward a contradiction, suppose that~$p$ still occurs in a data constraint~$\varphi$ in~$\bfa \subtry p$.
By the definition of~$\subtry$, we have~$\varphi = \existsfun_p(\varphi')$ for a data constraint~$\varphi'$ in~$\bfa$.
Because~$\existsfun$ does not introduce new variables in data constraints,~$p$ must have occurred already in~$\varphi'$.
Because~$p$ is an ever-determined port of~$\bfa$ by the premise of this theorem, by the definition of~$\edpfun$, we know that~$p$ has a determinant~$t$ in~$\varphi'$.
Consequently,~$\existsfun_p(\varphi') = \varphi' [t / p]$.
Also, from the fact that~$p$ has a determinant in~$\varphi'$, we can derive that~$p$ is not bound by any of the existential quantifications inside~$\varphi'$.
Hence,~$p$ does not occur in~$\varphi' [t / p]$.
But then,~$p$ does not occur in~$\existsfun_p(\varphi')$ either.
Therefore,~$p$ does not occur in~$\varphi$, which contradicts our intial assumption.
Hence,~$p$ does not occur in any data constraint in~$\bfa \subtry p$, which is the result stated in the consequence of this theorem.

A full, detailed proof appears as the proof of Theorem 15 in~\cite[Appendix D.3]{techreport}.
\qed

\subsection*{Proof of Theorem~\ref{thm:algo+commandification}}
\label{sect:proofs:thm+algo+commandification}

To show the correctness of Algorithm~\ref{algo:commandification} (henceforth ``the algorithm''), we need to show that if its \textbf{require}ments are satisfied, upon termination, it \textbf{ensure}s both:
$$
\LINES{\triplproves \tripl{\bigwedge \setb{x \termeq x}{x \in X}}{\pi}{\ell_1 \wedge \cdots \wedge \ell_i}}
$$
and
$$
\SCOPEX{%
	\sigma \dcmodels \ell_1 \wedge \cdots \wedge \ell_{n+m} \RIMPLIES 
\\	\UP
\\	\triplprovestot \triplx{\bigwedge \setb{x \termeq \sigma(x)}{x \in X}}{\pi}{\bigwedge \setb{x \termeq \sigma(x)}{x \in X \cup \set{x_1 \, \ldots \, x_n}}}
} \FORALL \sigma
$$
We call the former \emph{soundness} and the latter \emph{completeness} and prove their truth separately.
\begin{description}
	\item[Soundness]
	We start by arguing that~$\triplproves \tripl{\bigwedge \setb{x \termeq x}{x \in X}}{\pi}{\ell_1 \wedge \cdots \wedge \ell_i}$ holds after every iteration of the first loop.
	For~$1 \leq i \leq n$, after doing an assignment~$x_i \commasgn t_i$ in a data state~$\sigma$, literal~$\ell_i = x_i \termeq t_i$ holds in~$\sigma$ if all variables in~$t_i$ have a non-$\nil$ value.
	(Otherwise,~$t_i$ evaluates to~$\nil$, which the definition of~$\dcmodels$ forbids.)
	
	Reasoning toward a contradiction, suppose that some variable~$y$ in~$t_i$ has a~$\nil$ value.
	Then, because no assignment assigns~$\nil$, no~$y \commasgn t$ assignment has occurred previously.
	But because~$y \in \variablfun(t_i)$, either \SCOPE{a literal~$y \termeq t \in L$ exists that precedes~$x_i \termeq t_i$} or~$y \in X$ (by the \textbf{require}ments of the algorithm).
	In the former case, a~$y \commasgn t$ assignment must have occurred previously, such that~$y$ in fact has a non-$\nil$ value (namely, the evaluation of~$t$).
	In the latter case, by the precondition of the triple we are proving, we know that~$\sigma \dcmodels y \termeq y$ holds.
	By the definition of~$\dcmodels$, this means that~$y$ has a non-$\nil$ value.
	
	Thus,~$\ell_i = x_i \termeq t_i$ holds in~$\sigma$ after its update with~$x_i \commasgn t_i$.
	By the precondition of the triple, we know that~$x \termeq x$ held for all~$x \in X$ before updating~$\sigma$.
	Additionally, suppose that the preceding literals~$x_j \termeq t_j$ (for~$1 \leq j < i$) held before updating~$\sigma$.
	Each of those literals can have become false only if the update overwrote an~$x$ or an~$x_j$.
	In that case,~$x_i \in X \cup \set{x_1 , \ldots , x_{i-1}}$.
	But then, the algorithm did not translate~$x_i \termeq t_i$ to an assignment in the first place but to a failure statement~$\commifthen{x_i \termeq t_i}{\commskip}$.
	If execution of this statement successfully terminates, obviously~$x_i \termeq t_i$ holds, and because it leaves~$\sigma$ unchanged, all preceding literals remain true.
	Note that the~$\triplproves$ proof rule for failure statements allows us to \emph{assume} that the guard holds; we do not need to \emph{establish} this yet (cf. completeness below, where we use~$\triplprovestot$).
	
	We can inductively repeat the reasoning in the previous paragraphs for all~$1 \leq i \leq n$ to conclude that~$\triplproves \tripl{\bigwedge \setb{x \termeq x}{x \in X}}{\pi}{\ell_1 \wedge \cdots \wedge \ell_i}$ holds after the first loop.
	The failure statements added in the second loop leave state~$\sigma$ unchanged, meaning that literals that held before executing those statements in~$\sigma$ remain true.
	Thus, if those statements successfully terminate,
	$$\triplproves \tripl{\bigwedge \setb{x \termeq x}{x \in X}}{\pi}{\ell_1 \wedge \cdots \wedge \ell_{n+m}}$$ holds.
	
	\item[Completeness]
	Assume that~$\sigma' \dcmodels \ell_1 \wedge \cdots \wedge \ell_{n+m}$ for some~$\sigma'$.
	We start by arguing that~$\triplprovestot \tripl{\bigwedge \setb{x \termeq \sigma'(x)}{x \in X}}{\pi}{\bigwedge \setb{x \termeq \sigma'(x)}{x \in X \cup \set{x_1 \, \ldots \, x_i}}}$ holds after every iteration of the first loop.
	This means that the data state~$\sigma$ after executing~$\pi$ (starting from a data state where~$\bigwedge \setb{x \termeq \sigma'(x)}{x \in X}$ holds) maps every~$x_j$ (for~$1 \leq j \leq i$) to the same value as~$\sigma'$ (i.e.,~$\sigma(x_j) = \sigma'(x_j)$).
	Let~$1 \leq i \leq n$.
	
	If~$x_i \notin X \cup \set{x_1 , \ldots , x_{i-1}}$, we know that~$\ell_i = x_i \termeq t_i$ holds in~$\sigma$ after its update with~$x_i \commasgn t_i$ (see soundness above).
	By our initial assumption, we also know that~$\ell_i = x_i \termeq t_i$ holds in~$\sigma'$.
	Thus, by the definition of~$\dcmodels$, we conclude~$\sigma(x_i) = \evalfun_\sigma(t_i)$ and~$\sigma'(x_i) = \evalfun_{\sigma'}(t_i)$.
	Now, because a~$y \termeq t$ literal precedes~$x_i \termeq t_i$ for all~$y \in \variablfun(t_i)$ (see soundness above),~$\sigma$ maps every such a~$y$ to the same value as~$\sigma'$ (i.e.,~$y = x_j$ for some~$1 \leq j < i$).
	Consequently,~$\evalfun_\sigma(t_i) = \evalfun_{\sigma'}(t_i)$.
	Combining this with the previous intermediate result, the following equation holds:~$\sigma(x_i) = \evalfun_\sigma(t_i) = \evalfun_{\sigma'}(t_i) = \sigma'(x_i)$.
	Thus,~$x_i \termeq \sigma'(x_i)$ holds in~$\sigma$.
	As before (see soundness above), we can also establish that\editt{, for~$x_j \in X \cup \set{x_1 , \ldots , x_{i-1}}$,} updating~$\sigma$ with~$x_i \commasgn t_i$ does not make~$x_j \termeq \sigma'(x_j)$ literals that held already before this update false.
	Thus,~$\bigwedge \setb{x \termeq \sigma'(x)}{x \in X \cup \set{x_1 \, \ldots \, x_i}}$ holds in~$\sigma$.
	
	If~$x_i \in X \cup \set{x_1 , \ldots , x_{i-1}}$, we can immediately conclude that~$x_j \termeq \sigma'(x_j)$ held in~$\sigma$ for all~$x_j \in X \cup \set{x_1 , \ldots , x_{i-1}}$ already before executing the failure statement~$\commifthen{x_i \termeq t_i}{\commskip}$ added by the algorithm.
	To prove that this failure statement also successfully terminates, the~$\triplprovestot$ proof rule for failure statements dictates that we must establish---instead of assume (cf. soundness above)---that the guard~$x_i \termeq t_i$ holds in~$\sigma$.
	This follows from the fact that~$x_i \termeq t_i$ holds in~$\sigma'$ by our initial assumption, and because~$\sigma$ and~$\sigma'$ map all variables in~$\ell_i = x_i \termeq t_i$ to the same values.
	To prove the latter, we can use a similar argument involving the precedence relation and its linearization as before (see soundness above).
	
	We can inductively repeat the previous reasoning for all~$1 \leq i \leq n$ to conclude that~$\triplprovestot \tripl{\bigwedge \setb{x \termeq \sigma'(x)}{x \in X}}{\pi}{\bigwedge \setb{x \termeq \sigma'(x)}{x \in X \cup \set{x_1 \, \ldots \, x_n}}}$ holds after the first loop.
	The failure statements added in the second loop leave~$\sigma$
        unchanged,\break meaning that the~$x_j \termeq \sigma'(x_j)$
        literals that held already before executing those state- ments in~$\sigma$, for~$x_j \in X \cup \set{x_1 , \ldots , x_n}$, remain true.
	In order to prove the successful termination of those failure statements, we can use a similar argument as for the failure statements added in the first loop: by our initial assumption,~$\sigma' \dcmodels \ell_i$ for all~$n + 1 \leq j \leq n + m$, and~$\sigma$ and~$\sigma'$ still map the same variables to the same values.
	Thus,~$\triplprovestot \tripl{\bigwedge \setb{x \termeq \sigma'(x)}{x \in X}}{\pi}{\bigwedge \setb{x \termeq \sigma'(x)}{x \in X \cup \set{x_1 \, \ldots \, x_{n+m}}}}$ holds also after the second loop.
\end{description}

A full, detailed proof appears as the proof of Theorem 18 in~\cite[Appendix D.4]{techreport}.
\qed

\subsection*{Proof of Theorem~\ref{thm:algo+commandification+requir}}
\label{sect:proofs:thm+algo+commandification+requir}

Recall that the rules in Definition~\ref{def:prec} of~$\prec$ (and, therefore, also the rules in Definition~\ref{def:precx}) induce precedence relations for which all \textbf{require}ments of Algorithm~\ref{algo:commandification} (henceforth: ``the algorithm'') hold, except that those precedence relations \editt{do} not necessarily denote strict partial orders.
What we need to show here, then, is that~$\precstrict_\varphi^X$ is both a strict partial order and a ``large enough'' subset of~$\prec_\varphi^X$ to satisfy the algorithm's \textbf{require}ments.
The theorem subsequently follows, as~$\prectotal_\varphi^X$ is just the linearization of~$\precstrict_\varphi^X$.

The fact that~$\precstrict_\varphi^X$ is a strict partial order follows from~$\bprecarbor_\varphi^X$ forming an arborescence.

To show~${\precstrict_\varphi^X} \subseteq {\prec_\varphi^X}$, we need to consider the three rules in Definition~\ref{def:precstrict} of~$\precstrict$.
\emph{First}, take any pair~$\tpl{\ell \, \ell'}$ such that~$\ell \precstrict_\varphi^X \ell'$ by Rule \ref{rule:precstrict+free}.
Then, by the premise of that rule,~$\set{\ell_1 \, \ldots \, \ell_k} \bprecarbor_\varphi^X \ell'$ such that~$\ell = \ell_i$ for some~$1 \leq i \leq k$.
Because~${\bprecarbor_\varphi^X} \subseteq {\bprec_\varphi^X}$ (because the former is an arborescence of the latter), the premises of the rules in Definition~\ref{def:bprec} of~$\bprec$, subsequently guarantee after some manipulation that~$\ell = \ell_i = x \termeq t$ for some~$x$ and~$t$.
Moreover,~$x \in \variablfun(\ell')$.
By Rule \ref{rule:prec+free}, we subsequently conclude that~$\ell \prec_\varphi^X \ell'$ holds.
\emph{Second}, Rule~\ref{rule:precstrict+conv} is identical to Rule~\ref{rule:prec+conv}, so any pair~$\tpl{\ell , \ell'}$ in~$\precstrict_\varphi^X$ induced by the former is also induced in~$\prec_\varphi^X$ by the latter.
\emph{Third}, by induction, we can show the same result for pairs~$\tpl{\ell \, \ell'}$ such that~$\ell \precstrict_\varphi^X \ell'$ by Rule \ref{rule:precstrict+trans}.
Thus,~${\precstrict_\varphi^X} \subseteq {\prec_\varphi^X}$.

Finally, we must show that~${\precstrict_\varphi^X}$ is ``large enough'' for it to satisfy the precondition of the algorithm.
Informally, this means that arborescences do not exclude \barcs in the \bgraph that actually represent essential dependencies: for every free variable~$y$ that a literal~$\ell \in L$ depends on,~$\precstrict_\varphi^X$ must contain at least one pair~$\tpl{y \termeq t \, \ell}$ (for some~$t$).
To see that this holds, note that every \barc entering a literal~$\ell$ represents a complete set of dependencies of~$\ell$.
If~$\ell$ has multiple incoming \barcs, this simply means that several ways exist to resolve~$\ell$'s dependencies.
In principle, however, keeping one of those options suffices for our purpose.
Therefore, the single incoming \barc that~$\ell$ has in an arborescence represents enough dependencies of~$\ell$.

A full, detailed proof appears as the proof of Theorem 19 in~\cite[Appendix D.4]{techreport}.
\qed

\subsection*{Proof of Lemma~\ref{lemma:dcequiv+commfun}}
\label{sect:proofs:lemma+dcequiv+commfun}

Follows from Theorems~\ref{thm:algo+commandification} and~\ref{thm:algo+commandification+requir} and Definition~\ref{def:commfun}.

A full, detailed proof appears as the proof of Lemma 18 in~\cite[Appendix D.4]{techreport}.
\qed

\subsection*{Proof of Theorem~\ref{thm:congr+comm}}
\label{sect:proofs:thm+congr+comm}

Follows from Proposition~\ref{prop:001} and Lemma~\ref{lemma:dcequiv+commfun}.

A full, detailed proof appears as the proof of Theorem 20 in~\cite[Appendix D.4]{techreport}.
\qed

\subsection*{Proof of Theorem~\ref{thm:comm+effect}}
\label{sect:proofs:thm+comm+effect}

To prove this theorem, by Definition~\ref{def:comm} of~$\comm{\cdot}$, we need to show that for every data constraint~$\varphi$ in~$\bfa$, the pair~$\tpl{\varphi , X}$ for~$X = \freefun(\varphi) \cap (P^\textin \cup \pre{M})$ satisfies the four conditions in Definition~\ref{def:commfun} of~$\commfun$.
The first two conditions always hold.
The third condition follows from~$\arbor \bfa$: by Definition~\ref{def:arbor} of~$\arbor$, every data constraint in~$\bfa$ is arborescent.
Finally, the fourth condition follows from set theory.

A full, detailed proof appears as the proof of Theorem 21 in~\cite[Appendix D.4]{techreport}.
\qed	
\end{document}